\documentclass[preprintnumbers,unsortedaddress,superscriptaddress,notitlepage,square,sort&compress,comma,numbers]{revtex4-1}
\usepackage{lipsum}
\usepackage{CJK}
\usepackage{amsfonts}
\usepackage{amsmath}
\usepackage{amssymb}
\usepackage[english]{babel}
\usepackage{graphicx}
\usepackage{epsfig}
\usepackage{bm}
\usepackage{verbatim}
\usepackage[utf8]{inputenc}
\usepackage{booktabs}
\usepackage{multirow}
\usepackage{subfig}
\usepackage{slashed}
\usepackage{xcolor}
\usepackage[colorlinks=true,urlcolor=red,citecolor=red]{hyperref}
\usepackage[font=small]{caption}
\usepackage{float}
\usepackage{blindtext}
\usepackage{placeins}
\usepackage{bbm}
\usepackage[colorinlistoftodos]{todonotes}
\newcommand{\hs}{\hspace*{0.3cm}}
\newcommand{\cm}{\hspace*{1cm}}

\newcommand{\be}{\begin{equation}}
\newcommand{\ee}{\end{equation}}
\newcommand{\bea}{\begin{eqnarray}}
\newcommand{\eea}{\end{eqnarray}}
\newcommand{\ben}{\begin{enumerate}}
\newcommand{\een}{\end{enumerate}}
\newcommand{\bit}{\begin{itemize}}
\newcommand{\eit}{\end{itemize}}
\newcommand{\bde}{\begin{widetext}}
\newcommand{\ede}{\end{widetext}}
\newcommand{\nn}{\nonumber}
\newcommand{\crn}{\nonumber \\}

\newcommand{\al}{\alpha}
\newcommand{\la}{\lambda}
\newcommand{\bet}{\beta}
\newcommand{\ga}{\gamma}
\newcommand{\va}{\varphi}
\newcommand{\om}{\omega}
\newcommand{\pa}{\partial}
\newcommand{\+}{\dagger}
\newcommand{\fr}{\frac}

\newcommand{\bc}{\begin{center}}
\newcommand{\ec}{\end{center}}
\newcommand{\Ga}{\Gamma}
\newcommand{\de}{\delta}
\newcommand{\De}{\Delta}

\newcommand{\varep}{\varepsilon}

\newcommand{\La}{\Lambda}
\newcommand{\si}{\sigma}

\newcommand{\eq}{\eqref}

\newcommand{\mathsym}[1]{{}}

\input{tcilatex}
\newcommand{\Long}[1]{{#1}}
\newcommand{\Antonio}[1]{{#1}}

\newcommand{\crb}[1]{{\color{blue}#1}}

\newcommand{\AECH}[1]{{\color{blue}#1}}
\DeclareUnicodeCharacter{2212}{-}
\begin{document}

\title{An extended 3-3-1 model with two scalar triplets and linear seesaw mechanism}
\author{A. E. C\'arcamo Hern\'andez$^{a,b,c}$}
\email{antonio.carcamo@usm.cl}
\author{L.T. Hue$^{d,e}$}
\email{lethohue@duytan.edu.vn}
\author{Sergey Kovalenko$^{c,f}$}
\email{sergey.kovalenko@unab.cl}
\author{H. N. Long\footnote{Corresponding author}$^{ g,h}$}
\email{hoangngoclong@tdtu.edu.vn}
\affiliation{$^a$Universidad T\'{e}cnica Federico Santa Mar\'{\i}a, Casilla 110-V, Valpara\'{\i}so, Chile,\\
$^b$Centro Cient\'{\i}fico-Tecnol\'ogico de Valpara\'{\i}so, Casilla 110-V, Valpara\'{\i}so, Chile,\\
$^c$Millennium Institute for Subatomic Physics at High-Energy Frontier (SAPHIR), Fern\'andez Concha 700, Santiago, Chile\\
$^{d}$Institute for Research and Development, Duy Tan University, Da Nang City 55000, Vietnam\\
$^{e}$ Institute of Physics, Vietnam Academy of Science and Technology, 10 Dao Tan, Ba
Dinh, 100000 Hanoi, Vietnam\\
$^{f}$Departamento de Ciencias F\'isicas, Universidad Andres Bello, \\
Sazi\'e 2212, Piso 7, Santiago, Chile\\
$^{g}$Theoretical Particle Physics and Cosmology Research Group,\\
Advanced Institute of Materials Science, Ton Duc Thang University, Ho Chi
Minh City, Vietnam\\
$^{h}$ Faculty of Applied Sciences, Ton Duc Thang University,
Ho Chi Minh City, Vietnam}

\date{\today }

\begin{abstract}
Low energy linear seesaw mechanism responsible for the generation of the tiny active neutrino masses,
 is implemented in the extended 3-3-1 model with two scalar triplets and right handed Majorana neutrinos
  where the gauge symmetry is supplemented by the $A_4$ flavor discrete group and other auxiliary cyclic symmetries,
   whose spontaneous breaking produces the observed pattern of SM charged fermion masses and fermionic mixing parameters.
    Our model is consistent with the low energy SM fermion flavor data as well as with the constraints arising from meson oscillations.
Some phenomenological aspects such as the $Z^\prime$ production at proton-proton collider and the lepton
flavor violating decay of the SM-like Higgs boson are discussed. The scalar potential of the model is
 analyzed in detail and the SM-like Higgs boson is identified.

\end{abstract}

\maketitle

\allowdisplaybreaks
\section{Introduction}
\label{intro}

It is well-known, that there are various experimental and theoretical observations indicating
 that the Standard Model (SM) must be extended.
Among the theories beyond the SM, the models based on the gauge group
\mbox{$SU(3)_C\times SU(3) _L\times U(1) _X$} (called 3-3-1 for short)
\cite{Georgi:1978bv,Valle:1983dk,Pisano:1991ee,Foot:1992rh,Frampton:1992wt,Hoang:1996gi, Hoang:1995vq,Foot:1994ym,CarcamoHernandez:2005ka,Dong:2010zu,Dong:2010gk,Dong:2011vb,Benavides:2010zw,Dong:2012bf,Huong:2012pg,Giang:2012vs,Binh:2013axa,Hernandez:2013mcf,Hernandez:2013hea,Hernandez:2014vta,Hernandez:2014lpa,Kelso:2014qka,Vien:2014gza,Phong:2014ofa,Phong:2014yca,Boucenna:2014ela,DeConto:2015eia,Boucenna:2015zwa,Boucenna:2015pav,Benavides:2015afa,Hernandez:2015tna,Hue:2015fbb,Hernandez:2015ywg,Fonseca:2016tbn,Fonseca:2016xsy,Deppisch:2016jzl,Reig:2016ewy,CarcamoHernandez:2017cwi,CarcamoHernandez:2017kra,Hati:2017aez,Barreto:2017xix,CarcamoHernandez:2018iel,Vien:2018otl,Dias:2018ddy,Ferreira:2019qpf,Huong:2019vej,CarcamoHernandez:2019vih,CarcamoHernandez:2019iwh,CarcamoHernandez:2019lhv} have some intriguing features allowing them to explain the number of SM fermion families,
the electric charge quantization \cite{deSousaPires:1998jc,VanDong:2005ux},
etc.
In the ordinary 3-3-1 models, the Higgs sector contains at least three
scalar triplets significantly extending their scalar spectrum. Attempts aimed to reduce the Higgs sector of the 3-3-1 models have been undertaken in the literature.
A model with the parameter $\bet = - \fr 1 {\sqrt{3}}$, defined in (\ref{eq:Q-def}) and characterizing the embedding of the electric charge generator into $SU(3)_{L}$, has been proposed in
Refs.~\cite{Ponce:2002sg,Dong:2006mg,Dong:2006gx,Dong:2008ya,Ferreira:2011hm,Dong:2017ayu,CarcamoHernandez:2019vih}. Due to its restricted scalar sector it
is called the economical 3-3-1 model.
However, this and other similar versions of the 3-3-1 model with the reduced scalar content
failed to reproduce the neutrino oscillation data.
In a view of these difficulties a 3-3-1 model with $\bet = \fr 1 {\sqrt{3}}$ and containing just two Higgs triplets has been studied in Ref.~\cite{Barreto:2017xix}.
In this model the masses of light active neutrinos and charged fermions are generated via
Type-I Seesaw and the Universal Seesaw mechanisms, respectively. However, the fermion mixing was not addressed  in Ref.~\cite{Barreto:2017xix}.

In the present paper we propose a multiscalar singlet extension of the 3-3-1 model with two $SU(3)_L$ scalar triplets and three right handed Majorana neutrinos. The gauge group of the model is extended with the $A_4$ group and some other cyclic symmetries in order to implement the linear seesaw mechanism responsible for the tiny masses of the active neutrinos.  A well-known advantage of the linear seesaw mechanism \cite{Mohapatra:1986bd,Akhmedov:1995ip,Akhmedov:1995vm,Malinsky:2005bi,Borah:2018nvu,Hirsch:2009mx,Dib:2014fua,Chakraborty:2014hfa,Sinha:2015ooa,Borah:2018nvu,Das:2017ski} is
its testability at the LHC, since it implies sterile neutrinos with TeV-scale masses. Our model also successfully addresses the observed pattern of the SM fermion masses and mixings, as a result of the spontaneous breaking of the above mentioned discrete group factors, in an analogous way to the Froggat-Nielsen mechanism \cite{Froggatt:1978nt}, which has also been implemented in 3-3-1 models through the breaking of a $U(1)$ global symmetry in Refs.~\cite{Huitu:2017ukq,Huitu:2019kbm,Huitu:2019mdr}. We choose $A_4$ as the smallest discrete group having one three-dimensional and three distinct one-dimensional irreducible representations allowing us to naturally accommodate the three families of the SM. The $A_4$ discrete flavour group has received a lot interest by the model building community due to its remarkable ability to elucidate the observed pattern of SM fermion masses and mixing angles \cite{Ma:2001dn,He:2006dk,Feruglio:2008ht,Feruglio:2009hu,Chen:2009um,Varzielas:2010mp,Altarelli:2012bn,Ahn:2012tv,Memenga:2013vc,Felipe:2013vwa,Varzielas:2012ai, Ishimori:2012fg,King:2013hj,Hernandez:2013dta,Babu:2002dz,Altarelli:2005yx,Gupta:2011ct,Morisi:2013eca, Altarelli:2005yp,Kadosh:2010rm,Kadosh:2013nra,delAguila:2010vg,Campos:2014lla,Vien:2014pta,Joshipura:2015dsa,Hernandez:2015tna,Karmakar:2016cvb,Chattopadhyay:2017zvs,CarcamoHernandez:2017kra,Ma:2017moj,CentellesChulia:2017koy,Bjorkeroth:2017tsz,Srivastava:2017sno,Borah:2017dmk,Belyaev:2018vkl,CarcamoHernandez:2018aon,Srivastava:2018ser,delaVega:2018cnx,Borah:2018nvu,Pramanick:2019qpg,CarcamoHernandez:2019pmy,CarcamoHernandez:2019kjy,Ding:2019zxk,Okada:2019uoy}.

Comparing our model with others, we note, in particular, that our $U(1)_X$-charge assignments of the left handed quark $SU(3)_L$-triplets are different from those in the model of Ref. \cite{Barreto:2017xix}. Due to this difference  we have two exotic down type quarks and one exotic up type quarks whereas in the model of Ref. \cite{Barreto:2017xix} there are two exotic up type quarks and one exotic down type quark. In addition, whereas in our model the small masses for the active neutrinos are produced from a linear seesaw mechanism, in the model of Ref. \cite{Barreto:2017xix} they are generated from a type-I seesaw mechanism. In Ref. \cite{Barreto:2017xix}, the extra fermion lying in the bottom of the lepton triplet is a charged lepton instead of the
 right-handed neutrino, which is the field of the third component of $SU(3)_L$ leptonic triplet in our model.

Let us also note that
our model is more predictive and significantly more economical in its particle content than the 3-3-1 model with $T^{\prime}$ and $S_4$ symmetries proposed in \cite{CarcamoHernandez:2019vih,CarcamoHernandez:2019iwh}.
For instance, whereas the scalar sector of the $T^{\prime}$ flavored 3-3-1 model \cite{CarcamoHernandez:2019vih} includes two $SU(3)_L$ scalar triplets and $23$ gauge singlet scalar fields, the present model has two $SU(3)_L$ scalar triplets and $16$ $SU(3)_L$ singlet scalar fields. As for the scalar sector of the 3-3-1 model with $S_4$ family symmetry \cite{CarcamoHernandez:2019iwh}, it contains 3 $SU(3)_L$ scalar triplets and $32$ gauge singlet scalar fields, which is much larger than the number of scalar degrees of freedom of our model. Let us note, that in the proposed model some quarks and scalar fields carry lepton number, which
leads to flavor lepton number violating decay modes of the SM-like Higgs boson. In what follows we will study this phenomenological aspect of our model as well as the production of the extra heavy neutral gauge boson $Z'$ and its detection in the dimuon channel at the LHC. However, the emphasis will be made on studying the SM fermion masses and mixings.

The paper is organized as follows. In Sect. \ref{model} we introduce the model setup. Sects. \ref{quarksector} and \ref{leptonsector} are devoted to the model predictions for the masses and mixings in the quark and lepton sectors, respectively. Sect. \ref{FCNC} discusses the constraints on the $Z'$ mass arising from meson oscillations. In Sect. \ref{lfvdecay} the  lepton flavor violating (LFV) decays of the charged leptons and the Higgs boson are considered.  In Sect. \ref{conclusion} we summarize our results and discuss their further implications.  In Appendix \ref{A4} we present the discrete group $A_4$ group characters. A detailed description of  the Higgs sector of the model is given in Appendix  \ref{appHiggs}. The analytic formulas for one-loop contributions to the LFV decay amplitudes of the SM-like Higgs boson are collected in Appendix \ref{deltalr}. The couplings of neutral gauge bosons $Z$ and $Z'$ to fermions are listed in appendix~\ref{app_Zcoupling}.

\section{The model}
\label{model}

We propose a 3-3-1 model where the scalar sector is composed of two $%
SU( 3) _L$ scalar triplets and seven $SU( 3) _L$
scalar singlets and the fermion sector corresponds to the one of the 3-3-1
models with three right handed Majorana neutrinos. In our model the $%
SU(3)_C\times SU( 3) _L\times U( 1) _X$ \ gauge
symmetry is supplemented with the $A_4\times Z_8\times Z_{14}\times Z_{22}$
discrete group, so that the full symmetry $\mathcal{G}$ exhibits the
following three-step spontaneous breaking:%
\bea
&&\mathcal{G}=SU(3)_C\times SU(3)_L\times U(1)_X\times A_4\times
Z_8\times Z_{14}\times Z_{22}  \label{Group} \\
&&\hspace{35mm}\Downarrow \La _{int}  \crn[0.12in]
&&\hspace{15mm}SU(3)_C\times SU(3)_L\times U(1)_X  \crn[0.12in]
&&\hspace{35mm}\Downarrow v_\chi   \crn[0.12in]
&&\hspace{15mm}SU(3)_C\otimes SU( 2) _L\times U(
1) _Y  \crn[0.12in]
&&\hspace{35mm}\Downarrow v_\eta   \crn[0.12in]
&&\hspace{23mm}SU(3)_C\otimes U( 1) _Q  \nn
\eea
where the different symmetry breaking scales satisfy the following hierarchy
\be
v_\eta =v=246\mbox{GeV}\ll v_\chi \sim \mathcal{O}(10)\mbox{TeV} .
\label{eq:VEV-hierarchy}
\ee%
In the 3-3-1 model under consideration, the electric charge is defined in
terms of the $SU(3)$ generators and the identity by:
\be
\label{eq:Q-def}
Q=T_3+\beta T_8+X =T_3-\fr 1 {\sqrt{3}}T_8+X,
\ee%
where we have chosen $\beta =-\fr 1 {\sqrt{3}}$ (without non-SM electric charges), which implies that bottom
component of the lepton $SU(3)_L$-triplet is a neutral
field $\nu _R^C$ thus allowing to build the Dirac matrix with the usual
field $\nu _L$ in the top component of the lepton triplet. Adding gauge singlet right-handed Majorana neutrinos $N_{iR}$ $(i=1,2,3)$ will allow us
 to implement a low scale seesaw
mechanism, which could be inverse or linear, to generate the masses for the light active neutrinos.
 These low scale seesaw mechanisms offer attractive explanations for the smallest of neutrino masses because they can be tested at the LHC via the production and decay of sterile neutrinos. It is worth mentioning that the sterile neutrinos can be produced at the LHC in association
with a SM charged lepton and in pairs, via quark-antiquark annihilation mediated by a $W$ and heavy $W^\prime $ and $Z^\prime $ gauge bosons, respectively. In our model the sterile neutrinos have the following two body decay modes: $N^{\pm}_{a}\rightarrow l^{\pm}_iW^{\mp}$ and $N^{\pm}_{a}\rightarrow\nu_iZ$ (where $a,i=1,2,3$), which are suppressed by the small active-sterile neutrino mixing angle. Furthermore the heavy sterile neutrinos $N^{\pm}_{a}$ can decay via off-shell gauge bosons via the following modes: $N^{\pm}_{a}\rightarrow l^{+}_il^{-}_j\nu_k$, $%
N^{\pm}_{a}\rightarrow l^{-}_iu_j\bar{d}_k$, $N^{\pm}_{a}\rightarrow b\bar{b} \nu_k$ (where $a,i,j,k=1,2,3$ are flavour indices). Thus, the heavy sterile neutrinos can be detected at the LHC from  the observation of an excess of events with respect to the SM background in a final state composed of a pair of opposite sign charged leptons plus two jets. Studies of inverse seesaw neutrino signatures at colliders as well as the production of heavy neutrinos at the LHC are carried out in \cite{Dev:2009aw,BhupalDev:2012zg,Das:2012ze,AguilarSaavedra:2012fu,Das:2012ii,Dev:2013oxa,Das:2014jxa,Das:2016hof,Das:2017gke,Das:2017nvm,Das:2017zjc,Das:2017rsu,Das:2018usr,Das:2018hph,Bhardwaj:2018lma,Helo:2018rll,Pascoli:2018heg}. A detailed study of the sterile neutrino production at the LHC and the sterile neutrino modes goes beyond the scope of this work and will be done elsewhere.

The cancellation of chiral anomalies implies that the number of
triplets equals that of antitriplets, so that quarks are unified in the
following $SU(3)_C\times SU(3)_L\times U(1)_X$ left- and right-handed
representations \cite%
{Valle:1983dk,Hoang:1995vq,Diaz:2004fs,CarcamoHernandez:2005ka}:
\begin{align}
Q_{nL}& =\left(D_n  \, , -U_n  \, , J_n \, \right)^T_L \sim (
3,3^\ast ,0) ,\hspace{1cm}Q_{3L}=\left(U_3  \, , D_3  \, , \, T
\right)^T_L \sim \left( 3,3,\fr 1 3 \right) ,\cm n=1,2,  \crn
D_{iR}& \sim \left( 3,1,-\fr 1 3 \right) ,\cm U_{iR}\sim \left(
3,1,\fr 2 3 \right) ,\cm J_{nR}\sim \left( 3,1,-\fr 1 3 %
\right) ,\cm T_R \sim \left( 3,1,\fr 2 3\right) ,\quad
i=1,2,3 \, .\nn
\end{align}

Furthermore, the requirement of chiral anomaly cancellation constrains the
leptons to the following $SU(3)_C\times SU(3)_L\times U(1)_X$ left-
and right-handed representations \cite{Valle:1983dk,Hoang:1995vq,Diaz:2004fs}%
:
\be
L_{iL}=\left(\nu _i \, , e_i \, , \nu _i^c  \right)^T _L\sim \left(
1,3,-\fr 1 3 \right) ,\cm e_{iR}\sim ( 1,1,-1) ,%
\cm i=1,2,3,  \label{L}
\ee
 In the present model the fermion sector is extended by introducing three
right handed Majorana neutrinos, singlets under the 3-3-1 group, so that
they have the following $SU(3)_C\times SU(3)_L\times U(1)_X$
assignments:
\[
N_{iR}\sim ( 1,1,0) ,\cm i=1,2,3.
\]
Note that in the Ref.\cite{Barreto:2017xix}, where  $\beta =+\fr 1 {\sqrt{3}}$, the third component of lepton triplet is an
extra charged leptons.

We assign the scalar fields to the following $SU(3)_C\times
SU(3)_L\times U(1)_X$ representations:
\bea
\chi &=&%
\begin{pmatrix}
\chi _1 ^0 \\
\chi _2^- \\
\fr 1 {\sqrt{2}} (v_\chi +\xi _\chi \pm i\zeta _\chi )%
\end{pmatrix}%
\sim \left( 1,3,-\fr 1 3 \right) ,\cm \eta =%
\begin{pmatrix}
\fr 1 {\sqrt{2}} (v_\eta +\xi _\eta \pm i\zeta _\eta ) \\
\eta _2^- \\
\eta _3 ^0%
\end{pmatrix}%
\sim \left( 1,3,-\fr 1 3 \right) ,  \\
\label{eq:scalars-def}
\si &\sim &( 1,1,0) ,\cm \cm \xi _i\sim
( 1,1,0) ,\cm \cm \zeta _i\sim (
1,1,0) ,\cm \cm i=1,2,3.  \crn
\rho _i &\sim &( 1,1,0) ,\cm \cm \va
_i\sim ( 1,1,0) ,\cm \cm \phi _i\sim (
1,1,0) .
\nn
\eea
Here $v_{\chi}, v_{\eta}$ are the vev's setting symmetry breaking scales in (\ref{Group}), (\ref{eq:VEV-hierarchy}).\newline
 The scalar assignments under the \mbox{$A_4\times Z_8\times Z_{14}\times Z_{22}$} discrete group are summarized in Table \ref{tascalars}. 

In our model this discrete global symmetry group is not only spontaneously broken, it is softly broken as well. Let us note that the gauge singlet scalars of our models are complex, which implies that in order to provide masses for the CP odd parts of these scalars, one has to include $A_{4}\times Z_{8}\times Z_{14}\times Z_{22}$ soft breaking bilinear terms in the scalar potential involving a pair of these scalar singlets. These soft breaking scalar mass terms will also be useful for resolving the domain wall problem, arising from the spontaneous breaking the global discrete symmetries.

\begin{table}[th]
\begin{tabular}{|c|c|c|c|c|c|c|c|c|}
\hline
& $\chi$ & $\eta$ & $\si$ & $\xi$ & $\zeta$ & $\rho$ & $\phi$ & $\va$
\\ \hline
$A_4$ & $\mathbf{1}$ & $\mathbf{1}$ & $\mathbf{1}^{\prime}$ & $\mathbf{3}$
& $\mathbf{3}$ & $\mathbf{3}$ & $\mathbf{3}$ & $\mathbf{3}$ \\ \hline
$Z_8$ & $0$ & $-1$ & $0$ & $1$ & $-7$ & $-1$ & $-1$ & $4$ \\ \hline
$Z_{14}$ & $0$ & $-1$ & $0$ & $1$ & $-7$ & $1$ & $1$ & $1$ \\ \hline
$Z_{22}$ & $0$ & $-2$ & $-1$ & $2$ & $-1$ & $2$ & $2$ & $2$ \\ \hline
\end{tabular}%
\caption{Scalar assignments under $A_4\times Z_8\times Z_{14}\times
Z_{22}$.}
\label{tascalars}
\end{table}
In Appendix \ref{appHiggs} we present more details about the scalar sector of our model.

In what follows we briefly describe the gauge sector of our model.
Here we have 8 electroweak $SU(3)_L$ gauge bosons, $W_{a\mu }$, and a $U(1)_X$ gauge boson,
$\widetilde{B}_{\mu}$.

From the scalar kinetic term one finds the interactions:
 \bea
( D^\mu H)^\dag D_\mu H \supset \pa^\mu R_H^\+ P_\mu I_H - R^\+ P^\mu \pa_\mu I_H  \, \crb{,} \,  H= \eta, \chi ,
\label{c1}
\eea
The covariant derivative is defined as
\be
\label{eq:Covariant-Derivative}
D_{\mu }=\partial _{\mu }-iT_{a}W_{a\mu }-ig_{X}T_{9}X\widetilde{B}_{\mu
}=\partial _{\mu }-i\Pi _{\mu }
\ee
with:
\be
\Pi _{\mu }=\fr{g}{2}\left(
\begin{array}{ccc}
W_{3\mu }+\fr{1}{\sqrt{3}}W_{8\mu }+t\sqrt{\fr{2}{3}}X\widetilde{B}_{\mu
} & \sqrt{2}W_{\mu }^{+} & \sqrt{2}X_{\mu }^{q_{1}} \\
\sqrt{2}W_{\mu }^{-} & -W_{3\mu }+\fr{1}{\sqrt{3}}W_{8\mu }+t\sqrt{\fr{2%
}{3}}X\widetilde{B}_{\mu } & \sqrt{2}Y_{\mu }^{q_{2}} \\
\sqrt{2}X_{\mu }^{-q_{1}} & \sqrt{2}Y_{\mu }^{-q_{2}} & -\fr{2}{\sqrt{3}}%
W_{8\mu }+t\sqrt{\fr{2}{3}}X\widetilde{B}_{\mu }%
\end{array}%
\right) ,\hspace{1cm}t=\fr{g_{X}}{g}  \nn
\ee
where
\be
W_{\mu }^{\pm }=\fr{1}{\sqrt{2}}\left( W_{1}\mp iW_{2}\right) ,\hspace{1cm}%
X_{\mu }^{q_{1}}=\fr{1}{\sqrt{2}}\left( W_{4}-iW_{5}\right) ,\hspace{1cm}%
Y_{\mu }^{q_{2}}=\fr{1}{\sqrt{2}}\left( W_{6}-iW_{7}\right) .
\ee

Then, in the gauge sector we have three electrically neutral $q=0$ gauge fields, which combine to
 form the photon and $Z $, $Z^{\prime }$-bosons, two fields $W^{\pm }$ with $q=\pm 1$  and $X_{\mu }^{q_{1}}$, $Y_{\mu }^{q_{2}}$ with electrical charges
\bea
q_{1} &=&\fr{1}{2}+\fr{\sqrt{3}\beta }{2},\hspace{1cm}q_{2}=-\fr{1}{2}+\fr{\sqrt{3}\beta }{2}.
\eea
Physical neutral gauge bosons for $\beta=-\fr{1}{\sqrt{3}}$ are given by:
\bea
A_{\mu }&=&c_{W}\left( -\sqrt{\fr{1}{3}}t_{W}W_{8\mu }+\fr{\sqrt{%
3-4s_{W}^{2}}}{\sqrt{3}c_{W}}\widetilde{B}_{\mu }\right) +s_{W}W_{3\mu },\crn
Z_{\mu }&=&c_{W}W_{3\mu }-s_{W}\left( -\sqrt{\fr{1}{3}}t_{W}W_{8\mu }+\fr{%
\sqrt{3-4s_{W}^{2}}}{\sqrt{3}c_{W}}\widetilde{B}_{\mu }\right),\notag\\
Z_{\mu }^{\prime }&=&-\sqrt{\fr{1}{3}}t_{W}\widetilde{B}_{\mu }-\fr{\sqrt{%
3-4s_{W}^{2}}}{\sqrt{3}c_{W}}W_{8\mu }\crn
X_{\mu}^0&=&\fr{1}{\sqrt{2}}\left(W_{\mu 4}- iW_{\mu 5} \right),\hspace{1cm}\bar{X}_{\mu}^0=
\fr{1}{\sqrt{2}}\left(W_{\mu 4}+iW_{\mu 5} \right)
\label{eq:Neutral-Bosons}
\eea
where $c_W=\cos\theta_W$, $s_W=\cos\theta_W$ and $t_W=\tan\theta_W$, being $\theta_W$ the weak mixing angle.
In addition, for $\beta=-\fr1{\sqrt{3}}$, which corresponds to our model, we find the relations:
\be
\widetilde{B}_{\mu }=-\sqrt{\fr{1}{3}}t_{W}Z_{\mu }^{\prime }+\fr{\sqrt{%
3-4s_{W}^{2}}}{\sqrt{3}c_{W}}\left( c_{W}A_{\mu }-s_{W}Z_{\mu }\right) ,%
\hspace{1cm}W_{8\mu }=-\sqrt{\fr{1}{3}}t_{W}\left( c_{W}A_{\mu
}-s_{W}Z_{\mu }\right) -\fr{\sqrt{3-4s_{W}^{2}}}{\sqrt{3}c_{W}}Z_{\mu
}^{\prime },
\ee
\be
t=\fr{g_{X}}{g}=\fr{3\sqrt{2}s_{W}}{\sqrt{3-4s_{W}^{2}}}.
\ee
The electrically charged gauge bosons are given by:
\be
W_\mu^\pm = \fr 1 {\sqrt{2}} \left( A_{\mu 1} \mp i A_{\mu 2} \right)\, ,
\hs  Y_\mu^\pm = \fr 1 {\sqrt{2}} \left( A_{\mu 6} \pm i A_{\mu
7} \right)\,\label{eq9262}
\ee
where $Y^\pm$ and $X^0$ are bilepton gauge bosons.
With the above-discussed structure of the scalar sector of the model, the massive gauge bosons acquire the following masses
\cite{Long:2018dun}:
\be
m^2_W =m^2_Zc^2_W =\fr{g^2}{4}v_\eta^2 \, , \hs  M^2_{X^0}=M^2_{\bar{X}^0} = \fr{g^2}{4}%
\left(v^2_\chi +v_\eta^2\right) \, , \hs  M^2_Y = \fr{g^2}{4}%
v_\chi^2\, \, , \hs  M^2_{Z^\prime}\simeq \fr{g^2v_{\chi}^2}{3-t^2_W},
\label{eq9264}
\ee
where $v_\eta = v =246 $ GeV.
From (\ref{eq9264}) we find the mass splitting
\be
M^2_{X^0} - M^2_Y = m^2_W \, .  \label{eq9265}
\ee

In Ref.~\cite{Hoang:1999yv} it was shown
that the contributions of the bilepton gauge boson $Y^{\pm}, X^{0}$ to the oblique $S$ and $T$ parameters
 are constrained to be in the ranges $-0.085\lesssim S\lesssim 0.05$, $-0.001\lesssim T\lesssim 0.08$, respectively.
  In the scenario where the mixing angles between the exotic and the SM quarks are small, which is the the case of 
   our model, the exotic quark contributions to these oblique parameters are very subleading since they are 
   suppressed by the square of the small mixing angles. Consequently, the dominant contributions to the oblique
    $S$ and $T$ parameters are the ones arising from the bilepton gauge bosons $Y^\pm$ and $X^0$. 
    Notice that the aforementioned range of values for the $S$ and $T$ parameters allow one to have a region
     of the model parameter space where the obtained values for these oblique parameters are inside the 
     experimentally allowed region of Ref. \cite{Baak:2011ze} enclosed by the ellipses in the $S-T$ plane.

The fermion assignments under the $A_4\times Z_8\times Z_{14}\times
Z_{22}$ discrete group are summarized in Table \ref{ta:fermions}.

\begin{table}[th]
\begin{tabular}{|c|c|c|c|c|c|c|c|c|c|c|c|c|c|c|c|c|c|}
\hline
& $Q_{1L}$ & $Q_{2L}$ & $Q_{3L}$ & $U_{1R}$ & $U_{2R}$ & $U_{3R}$ & $T_R$ & $%
D_{1R}$ & $D_{2R}$ & $D_{3R}$ & $J_{1R}$ & $J_{2R}$ & $L_L$ & $N_R$ & $%
e_{1R} $ & $e_{2R}$ & $e_{3R}$ \\ \hline
$A_4$ & $\mathbf{1}^{\prime\prime}$ & $\mathbf{1}^\prime$ & $\mathbf{1}$
& $\mathbf{1}^{\prime\prime}$ & $\mathbf{1}^\prime$ & $\mathbf{1}$ & $%
\mathbf{1}$ & $\mathbf{1}^\prime$ & $\mathbf{1}^{\prime\prime}$ & $\mathbf{%
1}^{\prime\prime}$ & $\mathbf{1}^{\prime\prime}$ & $\mathbf{1}^\prime$ & $%
\mathbf{3}$ & $\mathbf{3}$ & $\mathbf{1}$ & $\mathbf{1}^\prime$ & $\mathbf{%
1}$ \\ \hline
$Z_8$ & $0$ & $0$ & $0$ & $-3$ & $-1$ & $1$ & $0$ & $2$ & $4$ & $-1$ & $0$ & $%
0$ & $0$ & $0$ & $0$ & $3$ & $0$ \\ \hline
$Z_{14}$ & $0$ & $0$ & $0$ & $5$ & $-1$ & $1$ & $0$ & $2$ & $4$ & $-1$ & $0$
& $0$ & $-4$ & $-4$ & $-6$ & $-6$ & $-6$ \\ \hline
$Z_{22}$ & $-5$ & $-4$ & $0$ & $-8$ & $-6$ & $2$ & $0$ & $-14$ & $-9$ & $-1$
& $-5$ & $-4$ & $-7$ & $-8$ & $-5$ & $-9$ & $-11$ \\ \hline
\end{tabular}%
\caption{Fermion assignments under $A_4\times Z_8\times Z_{14}\times
Z_{22}$.}
\label{ta:fermions}
\end{table}

We assume the following VEV pattern for the $A_4$ triplet SM singlet
scalars $\xi $, $\zeta $, $\rho $, $\va $ and $\phi $:%
\bea
\left\langle \xi \right\rangle &=&\fr{v_\xi  }{\sqrt{3}}(
1,1,1) ,\cm \left\langle \zeta \right\rangle =\fr{v_\zeta %
}{\sqrt{2}}( 1,0,1) ,\cm \left\langle \rho \right\rangle =%
\fr{v_\rho }{\sqrt{3}}( 1,1,1) ,  \crn
\left\langle \va \right\rangle &=&\fr{v_\va }{\sqrt{3}}\left(
\cos \al +e^{i\psi}\sin \al ,\om \left( \cos \al +\om e^{i\psi}\sin \al
\right) ,\om ^2\left(\cos \al +\om ^2e^{i\psi}\sin \al \right)
\right) ,\label{VEVpattern} \\
\left\langle \phi \right\rangle &=&\fr{v_\phi }{\sqrt{3}}\left( \cos
\al -e^{-i\psi}\sin \al ,\om ^2\left( \cos \al -\om ^2e^{-i\psi}\sin \al
\right) ,\om \left( \cos \al -\om e^{-i\psi}\sin \al \right) \right) ,%
\cm \om =e^{\fr{2\pi i}{3}},\nn
\eea%
which are consistent with the scalar potential minimization equations for a
large region of parameter space, as shown in details in Refs. \cite%
{Hernandez:2016eod,CarcamoHernandez:2017kra}.

With the above particle content, the relevant Yukawa terms for the quark and
lepton sectors invariant under the group $\mathcal{G}$ are:
\bea
-\mathcal{L}_Y^{( q) } &=&y^{( T) }\overline{Q}%
_{3L}\chi T_R +y_{33}^{( U) }\overline{Q}_{3L}\eta
U_{3R}\crn
&& +y_{22}^{( U) }\varep _{abc}\overline{Q}_{2L}^a\eta
^b \chi ^c U_{2R}\fr{( \xi \xi ) _{\mathbf{\mathbf{1}}}}{%
\La ^3 }+y_{11}^{( U) }\varep _{abc}\overline{Q}%
_{1L}^a \eta ^b\chi ^c U_{1R}\fr{\left( \xi ^3 \zeta \right) _{%
\mathbf{\mathbf{1}}}}{\La^{5}}  \crn
&&+y_1^{( J) }\overline{Q}_{1L}\chi ^\ast
J_{1R}+y_2^{( J) }\overline{Q}_{2L}\chi ^\ast
J_{2R}\crn
&& +y_{33}^{( D) }\varep _{abc}\overline{Q}%
_{3L}^a\left( \eta^\ast \right)^b\left( \chi ^\ast \right)
^c D_{3R}\fr{\si }{\La^2} +y_{22}^{( D) }\overline{Q}%
_{2L}\eta ^\ast  D_{2R}\fr{\left( \xi ^2\zeta \right) _{\mathbf{\mathbf{%
1}}}}{\La^3 }  \label{Lyq} \\
&&+y_{11}^{( D) }\overline{Q}_{1L}\eta ^\ast D_{1R}\fr{\left(
\xi ^4 \zeta \right) _{\mathbf{\mathbf{1}}}}{\La^5}+y_{12}^{(
D) }\overline{Q}_{1L}\eta ^\ast D_{2R}\fr{\left( \xi ^2\zeta
\right) _{\mathbf{\mathbf{1}}}\si }{\La^4}+y_{13}^{( D)
}\overline{Q}_{1L}\eta ^\ast D_{3R}\fr{\si ^6}{\La ^6}%
+y_{23}^{( D) }\overline{Q}_{2L}\eta ^\ast D_{3R}\fr{\si
^5}{\La^5}+H.c,  \nn
\eea%
\bea
-\mathcal{L}_Y^{( l) } &=&y_1 ^{( L) }\varep
_{abc}\left( \overline{L}_L^a\left( \eta ^\ast \right) ^b\left( \chi
^\ast \right) ^c\rho \right) _{\mathbf{\mathbf{1}}}e_{1R}\fr{\si
^6}{\Antonio{\La^8}}+y_2^{( L) }\varep _{abc}\left(
\overline{L}_L^a\left( \eta ^\ast \right) ^b\left( \chi ^\ast
\right) ^c \va \right) _{\mathbf{\mathbf{1}}}e_{2R}\fr{\si ^2}{%
\La^4}+\fr{y_3 ^{( L) }}{\La ^2}\varep
_{abc}\left( \overline{L}_L^a \left( \eta ^\ast \right) ^b\left( \chi
^\ast \right) ^c\rho \right) _{\mathbf{\mathbf{1}}}e_{3R}\crn
&&+z_1^{(L) }\varep _{abc}\left( \overline{L}%
_L^a \left( \eta^\ast \right) ^b\left( \chi^\ast \right)^c \phi
\right) _{\mathbf{\mathbf{1}}}e_{1R}\fr{\si^6}{\Antonio{\La^8}}+\fr{%
z_3 ^{( L) }}{\La ^2}\varep _{abc}\left( \overline{L}%
_L^a \left( \eta^\ast \right)^b\left( \chi^\ast \right)^c \phi
\right) _{\mathbf{\mathbf{1}}}e_{3R} \crn
&&+y^{(L)}_\rho\varep _{abc}\varep _{dec}\left( \overline{L}%
_L^a\left( L_L^C\right)^b\right) _{\mathbf{3a}}\eta^d\chi^e%
\fr{\zeta \si ^{11}}{\La ^{13}}+y_{1\eta }^{( L) }\left(
\overline{L}_L\eta N_R\right) _{\mathbf{3s}}\fr{\xi \si^\ast }{%
\La ^2}+y_{2\eta }^{( L) }\left( \overline{L}_L\eta
N_R \right) _{\mathbf{3a}}\fr{\xi \si^\ast }{\La ^2}\crn
&&+y_\chi ^{( L) }\left( \overline{L}_L\chi N_R\right) _{%
\mathbf{1}^\prime }\fr{\si^\ast } \La +H.c.\ ,  \label{Lyl}
\eea
where the dimensionless couplings $y, z$ in Eqs. \eq{Lyq} and  \eq{Lyl} are
$\mathcal{O}(1)$ parameters.
In addition to these terms, the symmetries unavoidably allow the following terms:
\bea
&&y^{( T3) }\overline{Q}_{3L}\chi U_{3R} \fr{(\xi^{*2}\xi)_1}{\La^3},
\hspace{2cm}y^{( 3T) }\overline{Q}_{3L}\eta T_{R} \fr {(\xi^*\xi^2)_1}{\La^3}\crn
&&y^{(J_2D_3)}\overline{Q}_{2L}\chi ^\ast D_{3R} \fr {(\xi^2\xi^*)_1\si^5}{\La^8},
\hspace{2cm}y^{( D_3J_2) }\varep _{abc}\overline{Q}%
_{3L}^a\left( \eta ^\ast \right) ^b\left( \chi ^\ast \right)
^c D_{3R}\fr{(\xi^{\ast2}\xi)_{1'}\si^{\ast4} }{\La^8}\, .
\nn
\eea
These terms will generate very small mixing angles of the third generation SM up and down type quarks with
 the exotic quarks. Such mixing angles are of the order of $\la^5$ and $\la^{11}$ (being $\la=0.225$), for the
  up and down type quarks, respectively, thus allowing us to safely neglect these strongly suppressed corrections, 
  which will not be considered in our analysis. Furthermore, as it will shown in Sect. \ref%
{quarksector}, the quark assignments under the different group factors of
our model will give rise to SM quark mass textures where the CKM quark
mixing angles only arise from the down type quark sector. As
indicated by the current low energy quark flavor data encoded in the
standard parametrization of the quark mixing matrix, the complex phase
responsible for CP violation in the quark sector is associated with the
quark mixing angle in the $1$-$3$ plane. Thus, the Yukawa coupling $%
y_{13}^{( D) }$ in Eq. \eq{Lyq} is required to be complex in
order to successfully reproduce the experimental values of the
quark mixing angles and CP violating phase.

In a generic scenario the Yukawa couplings are complex. However, not all of
them are physical. Some phases can be rotated away by the phase rotation of
the quark and lepton fields. The conditions for the rotation away of the
Yukawa phases in the quark sector by the redefinition of the phases $\alpha
_{f}$ of the quark fields are: 
\bea 
\arg \left( y_{33}^{\left( D\right) }\right) -\alpha _{Q_{3L}}+\alpha
_{D_{3R}} &=&0,\hspace{1cm}\arg \left( y_{23}^{\left( D\right) }\right)
-\alpha _{Q_{2L}}+\alpha _{D_{3R}}=0, \crn 
\arg \left( y_{13}^{\left( D\right) }\right) -\alpha _{Q_{1L}}+\alpha
_{D_{3R}} &=&0,\hspace{1cm}\arg \left( y_{22}^{\left( D\right) }\right)
-\alpha _{Q_{2L}}+\alpha _{D_{2R}}=0,  \crn 
\arg \left( y_{12}^{\left( D\right) }\right) -\alpha _{Q_{1L}}+\alpha
_{D_{2R}} &=&0,\hspace{1cm}\arg \left( y_{11}^{\left( D\right) }\right)
-\alpha _{Q_{1L}}+\alpha _{D_{1R}}=0, \crn 
\arg \left( y_{1}^{\left( J\right) }\right) -\alpha _{Q_{1L}}+\alpha
_{J_{1R}} &=&0,\hspace{1cm}\arg \left( y_{2}^{\left( J\right) }\right)
-\alpha _{Q_{2L}}+\alpha _{J_{2R}}=0, \crn 
\arg \left( y_{11}^{\left( U\right) }\right) -\alpha _{Q_{1L}}+\alpha
_{U_{1R}} &=&0,\hspace{1cm}\arg \left( y_{22}^{\left( U\right) }\right)
-\alpha _{Q_{2L}}+\alpha _{U_{2R}}=0, \crn 
\arg \left( y_{33}^{\left( U\right) }\right) -\alpha _{Q_{3L}}+\alpha
_{U_{3R}} &=&0,\hspace{1cm}\arg \left( y^{\left( T\right) }\right) -\alpha
_{Q_{3L}}+\alpha _{T_{R}}=0,
\eea %
Consequently all the Yukawa phases in the quark sector can be rotated away,
unless one considers phases of the scalar fields. Therefore, without
considering phase rotation of the scalar fields, all the Yukawa couplings of
the quark sector can be set real. Thus, in view of the above, the observed CP violation in
the quark sector will arise from complex vacuum expectation values of the
gauge singlet scalars charged under the discrete symmetries of the model.
Therefore, the spontaneous breaking of the discrete symmetries of our model,
gives rise to the observed CP violation in the quark sector. This mechanism
of generating CP violation in the fermion sector from the spontaneous
breaking of the discrete groups is called Geometrical CP violation and has
been implemented in other models. A concise review of group theoretical origin of CP violation is provided
in Ref.~\cite{Chen:2019iup}

Next, we explain the reason for introducing the discrete group factors in our model. We introduce the $A_4$ and $Z_{14}$ 
discrete groups  with the aim of reducing the number of model parameters, thus making our model more predictive. 
In addition, these discrete groups allow us to
get predictive and viable textures for the fermion sector capable of successfully explaining
the observed pattern of fermion
masses and mixing angles, as will be shown in Sects. \ref{quarksector} and %
\ref{leptonsector}. The $A_4$ and $Z_{14}$ discrete groups select the
allowed entries of the mass matrices for SM quarks.

The $Z_8$ discrete symmetry separates the $A_4$ scalar triplet $\xi $ participating in the charged
lepton Yukawa interactions from the remaining $A_4$ scalar triplets. The $%
Z_{14}$ discrete symmetry separates the $A_4$ scalar triplet
$\zeta $ participating in the Dirac neutrino Yukawa interactions from the $%
A_4$ scalar triplet \ $\xi $ appearing in some of the neutrino Yukawa
interactions involving the right handed Majorana neutrinos $N_{iR}$ ($i=1,2,3
$). Let us note that the different $A_4\times Z_{14}\times Z_{22}$ charge
assignments for the quark fields shown in Table \ref{ta:fermions} give rise to a CKM
quark mixing matrix solely emerging from the down type quark sector. The spontaneous breaking of
the $Z_{14}\times Z_{22}$ discrete group yields the hierarchical structure of
the SM charged fermion mass matrix and quark mixing angles.
Furthermore, the $Z_{22}$ symmetry is the
smallest cyclic symmetry allowing one to construct a Dirac Yukawa term
$\left(\overline{L}_L^a\left( L_L^C\right) ^b\right) _{\mathbf{3a}}\eta
^d\chi ^e\fr{\zeta \si ^{11}}{\La ^{13}}$ of dimension thirteen
from an $\fr{\si ^{11}}{\La ^{11}}$ insertion on the $\left(
\overline{L}_L^a\left( L_L^C\right) ^b\right) _{\mathbf{3a}}\eta
^d\chi ^e\fr{\zeta }{\La ^2}$ operator, necessary for obtaining the
required $\la ^{19}$ suppression (where $\la =0.225$ is one of the
Wolfenstein parameters) crucial for natural explanation of the smallness of the Dirac
neutrino mass matrix and thus of the light active neutrino masses, as it
will be explained in more details  in Sect. \ref{leptonsector}. Thus, in view of the above, the hierarchy
among charged fermion
 masses and quark mixing angles is caused by the spontaneous breaking of the
  $A_4\times Z_{14}\times Z_{22}$ discrete group.
Consequently, the quark masses are related with the quark
mixing angles and we therefore set the VEVs of the scalar fields $\eta $, $\chi $, $\si $%
, $\xi _j$, $\zeta _j$ ($j=1,2,3$) with respect to the
Wolfenstein parameter $\la $ and the model cutoff $\La $, as
follows:
\be
v_\eta \sim \la ^4\La < v_\zeta \sim \la ^3 \La
< v_\chi \sim \la ^2\La < v_\xi  \sim v_\si \sim v_\rho
\sim v_\va \sim v_\phi \sim \la \La \,.  \label{VEVsinglets}
\ee
It is worth mentioning, as follows from Eqs. (\ref{Lyq}) and (\ref{Lyl}) that the Yukawa interactions have a total of 21 parameters 
from which 18 are assumed to be real and 3 are taken to be complex. However not all of these parameters enter in the 
physical observables of the quark and lepton sectors. Such physical observables are determined by the resulting low energy 
SM fermion mass matrices which do depend on effective parameters which contain some of the Yukawa couplings as well 
as the VEVs of the scalar fields of the model. After the assumption shown in Eq. (\ref{VEVsinglets}) is made and the 
benchmarks described in sections III and IV are considered, the number of effective parameters can be reduced.
	
Furthermore, the VEV hierarchy $v_{\eta} \ll v_{\chi}\sim v_\zeta  \ll  v_\xi  \sim v_\si \sim v_\rho \sim v_\va \sim v_\phi$ is 
followed from the SSB chain of Eq. (1) and it also follows from gauge boson mass expressions: for example,
masses of the SM gauge bosons depend on $v_\eta$ while masses of new gauge bosons (X,Y) and $Z'$ depend 
on $v_\chi$.  In addition, the VEV hierarchy $v_\zeta  \ll  v_\xi  \sim v_\si \sim v_\rho \sim v_\va  \sim v_\phi$ 
can be explained by an appropriate relations between the different mass coefficients of the bilinear terms of 
the scalar potential and the VEVs of such scalar fields.
This can be explicitly shown by considering the simplified scenario of two singlet scalar fields $S_1$ and $S_2$, 
whose VEVs satisfy the hierarchy $v_{S_2} \ll v_{S_1}$. The scalar potential for such singlet fields is:
\be
V=-\mu _{S_1}^2\left\vert S_1\right\vert ^2-\mu
_{S_{2}}^{2}\left\vert S_{2}\right\vert ^{2}+\la  _{1}\left\vert
S_{1}\right\vert ^{4}+\la  _{2}\left\vert S_{2}\right\vert ^{4}+\la
_{3}\left\vert S_{1}\right\vert ^{2}\left\vert S_{2}\right\vert ^{2}.
\ee
Its minimization implies:
\be
\mu _{S_{1}}^{2}=2\la  _{1}v_{S_{1}}^{2}+\la  _{3}v_{S_{2}}^{2},%
\hspace{1.5cm}\mu _{S_{2}}^{2}=2\la  _{2}v_{S_{2}}^{2}+\la
_{3}v_{S_{1}}^{2}.
\ee
Thus, the VEV hierarchy $v_{S_{2}} \ll v_{S_{1}}$, can be justified by requiring $\mu _{S_{2}}^{2}\simeq 2\mu _{S_{1}}^{2}$ 
and considering the case where the quartic scalar couplings satisfy $\la_i\simeq\la$ ($i=1,2,3$). A straightforward but tedious 
extension of the aforementioned argument will yield to a large set of relationships between the different mass coefficients
 of the bilinear terms of the scalar potential and the VEVs of the large number of gauge singlet scalar fields of our model
  that will generate the VEV hierarchy shown in Eq. (\ref{VEVsinglets}).
	
It is worth mentioning that there are several operators invariant under the $%
SU\left( 3\right) _{C}\times SU\left( 3\right) _{L}\times U\left( 1\right)
_{X}$ gauge symmetry that can generate flavour and/or baryon
number violation. Following \cite{Grzadkowski:2010es}, we find that these operators are given by:

\bea 
&&\overline{L}_{iL}L_{jL}^{C}L_{kL}\overline{L_{rL}^{C}},
\hspace{1cm}
\overline{L}_{iL}e_{jR}\overline{Q}_{nL}U_{kR},\hspace{1cm}
\overline{Q}_{nL}Q_{mL}^{C}
\overline{Q}_{3L}L_{iL}^{C},\hspace{1cm}
\overline{Q}_{nL}L_{iL}^{C}U_{iR}^{C}\overline{D}_{jR},  \crn 
&&\overline{L}_{iL}e_{jR}Q_{3L}\overline{D}_{kR},
\hspace{1cm}
\overline{Q}_{nL}Q_{3L}^{C}\overline{U}_{jR}e_{kR}^{C},\hspace{1cm}
\overline{D}_{iR}U_{jR}^{C}%
\overline{U}_{kR}e_{rR}^{C},\hspace{1cm}
\overline{Q}_{nL}Q_{3L}^{C}U_{iR}\overline{D_{jR}^{C}},  \crn 
&&\overline{L}_{iL}Q_{3L}^{C}Q_{3L}\overline{L_{jL}^{C}
},\hspace{1cm}
\overline{Q}_{3L}Q_{mL}^{C}Q_{3L}
\overline{Q_{sL}^{C}},\hspace{1cm}
\overline{Q}_{nL}Q_{mL}^{C}Q_{pL}\overline{Q_{sL}^{C}},\hspace{1cm}
\overline{Q}_{3L}Q_{3L}^{C}Q_{3L}\overline{Q_{3L}^{C}}.  \label{op}
\eea 
where all subindices go from $1$ to $3$ excepting $n$, $m$, $s$ and $p$,
which take the values of $1$ and $2$. However all these operators, excepting 
$\overline{Q}_{3L}Q_{3L}^{C}Q_{3L}\overline{Q_{3L}^{C}}
$, are forbidden by the $A_{4}\times Z_{8}\times Z_{14}\times Z_{22}$
discrete symmetry. 
Despite this operator contributes to proton decay, it is phenomenologically innocent, since 
its contribution is suppressed by the eight power
of the very small $\theta _{13}^{\left( q\right) }\sim \lambda ^{4}$ ($%
\lambda =0.225$) quark mixing angle.
	
\section{Quark masses and mixings}
\label{quarksector}
From the quark Yukawa interactions given by Eq. \eq{Lyq} we find the following expressions for the non-vanishing elements of the SM up and down quark mass matrices
\bea 
&&M_{U11}=y_{11}^{\left( U\right) }\frac{v_{\chi
	}\left\langle \left( \xi ^{3}\zeta \right) _{\mathbf{1}}\right\rangle }{\sqrt{2}\Lambda ^{5}} \fr{v}{\sqrt{2}},\hspace{1cm}
M_{U22}=y_{22}^{\left(
	U\right) }\frac{v_{\chi }\left\langle \left( \xi \xi \right) _{\mathbf{1}}\right\rangle }{\sqrt{2}\Lambda ^{3}} \fr{v}{\sqrt{2}},\hspace{1cm}
M_{U33}=y_{33}^{\left( U\right) }  \fr{v}{\sqrt{2}},\crn 
&&M_{D11}=y_{11}^{\left( D\right) }\frac{\left\langle \left( \xi ^{4}\zeta \right) _{\mathbf{1}}\right\rangle }{\Lambda ^{5}} \fr{v}{\sqrt{2}},  
\hspace{1cm} 
M_{D12}=y_{12}^{\left( D\right) }\frac{v_{\sigma
	}\left\langle \left( \xi ^{2}\zeta \right) _{\mathbf{1}}\right\rangle }{\La^{4}} \fr{v}{\sqrt{2}},\hspace{1cm}
M_{D13}=y_{13}^{\left( D\right) }\frac{v_{\sigma }^{6}}{\Lambda ^{6}} \fr{v}{\sqrt{2}},
\crn
&&M_{D22}=y_{22}^{\left( D\right) }\frac{\left\langle \left( \xi ^{2}\zeta \right) _{\mathbf{1}}\right\rangle }{\Lambda ^{3}} \fr{v}{\sqrt{2}},\hspace{1cm} 
M_{D23}=y_{23}^{\left( D\right) }\frac{v_{\sigma }^{5}}{\Lambda ^{5}} \fr{v}{\sqrt{2}},  
\hspace{1cm} M_{D33}=-y_{33}^{\left( D\right) }\frac{v_{\chi
	}v_{\sigma }}{\sqrt{2}\Lambda ^{2}} \fr{v}{\sqrt{2}}.
\eea 
where $v= 246$ GeV is the scale of electroweak symmetry breaking and $\left\langle....\right\rangle$ stands for the vacuum expectation value of the product of the singlet scalar fields. 
For the VEV pattern of our model (\ref{VEVsinglets}) we find for the SM quark mass matrices:
\be
M_U=\left(
\begin{array}{ccc}
	a_1 ^{( U) }\la^8 & 0 & 0 \\
	0 & a_2^{( U) }\la^4 & 0 \\
	0 & 0 & a_3 ^{( U) }%
\end{array}%
\right) \fr{v}{\sqrt{2}},\cm \cm %
M_D=\left(
\begin{array}{ccc}
	a_{11}^{( D) }\la^7 & a_{12}^{( D) }\la^6
	& a_{13}^{( D) }\la^6 \\
	0 & a_{22}^{( D) }\la^5 & a_{23}^{( D) }\la^5 \\
	0 & 0 & a_{33}^{( D) }\la^3 %
\end{array}%
\right) \fr{v}{\sqrt{2}},  \label{Mq}
\ee
where $a_1 ^{( U) }, a_{11}^{( D) }, ...$
are $\mathcal{O}(1)$ dimensionless parameters being products of the dimensionless
couplings $y^{(K)}$ in Eq.~(\ref{Lyq}). 

Note that due to different $A_4\times Z_{14}\times Z_{22}$ charge assignments of  the quark fields, the exotic and the SM quarks do not mix with each other. Thus the exotic quark masses are:
\be
m_T =y^{( T) }\fr{v_\chi }{\sqrt{2}},\cm %
m_{J^1 }=y_1 ^{( J) }\fr{v_\chi }{\sqrt{2}}=\fr{%
y_1 ^{( J) }}{y^{( T) }}m_T ,\cm %
m_{J^2}=y_2^{( J) }\fr{v_\chi }{\sqrt{2}}=\fr{%
y_2^{( J) }}{y^{( T) }}m_T .  \label{mexotics}
\ee

As seen from Eq.~(\ref{Mq}), the model has ten physical parameters, allowing one reproduce any value
 of ten observables: six quark masses, three mixing angles and one Jarlskog CP invariant shown in Table \ref{Tab}. 
 The corresponding values of the model parameters are:
\be
\begin{array}{c}
a_1^{( U) }\simeq 1.085\,,\cm a_2^{( U)
}\simeq 1.391\,,\cm a_3 ^{( U) }\simeq 0.994\,, \\
a_{11}^{( D) }\simeq 0.527\,,\cm a_{22}^{( D)
}\simeq 0.491\,,\cm a_{33}^{( D) }\simeq 1.438\,, \\
a_{12}^{( D) }\simeq 0.501\,,\cm \left\vert
a_{13}^{( D) }\right\vert \simeq 0.467\,,\cm \arg
(a_{13}^{( D) })\simeq -60.96^{\circ }\,,\cm a_{23}^{(
D) }\simeq 1.210\,.%
\end{array}
\label{eq:bm-values}
\ee
An important feature of the above result is that the absolute values of all $a$-parameters are of the order of unity. 
Thus, the symmetries of our model allow us to naturally explain the hierarchy of quark mass spectrum without
 appreciable tuning of  these effective parameters.
\begin{table}[th]
\begin{center}
\begin{tabular}{c|l|l|l|l|l}
\hline
Observable & S-4 & S-3 & S-2a & S-2b & Experimental value \\ \hline
$m_{u}(\mathrm{MeV})$ & \quad $1.12$ & \quad $1.12$ & \quad $1.12$ & \quad $1.12$ & \quad $1.24\pm 0.22$ \\ \hline
$m_c(\mathrm{GeV})$ & \quad $0.617$ & \quad $0.617$  & \quad $0.617$  & \quad $0.617$ & \quad $0.63\pm 0.02$ \\ \hline
$m_t (\mathrm{GeV})$ & \quad $174$ & \quad $174$ & \quad $174$ & \quad $174$ & \quad $172.9\pm 0.4$ \\ \hline
$m_d (\mathrm{MeV})$ & \quad $2.64$ & \quad $2.23$ & \quad $2.24$ & \quad $2.34$ & \quad $2.69\pm 0.19$ \\ \hline
$m_s (\mathrm{MeV})$ & \quad $54.7$ & \quad $46.4$ & \quad $46.4$ & \quad $48.5$ & \quad $53.5\pm 4.6$ \\ \hline
$m_b (\mathrm{GeV})$ & \quad $2.76$ & \quad $2.76$ & \quad $2.76$ & \quad $2.76$ & \quad $2.86\pm 0.03$ \\ \hline
$\sin \theta^{(q)}_{12}$ & \quad $0.220$ & \quad $0.220$ & \quad $0.220$ & \quad $0.220$ & \quad $0.2245\pm 0.00044$ \\ \hline
$\sin \theta^{(q)}_{23}$ & \quad $0.0506$ & \quad $0.0506$ & \quad $0.0506$ & \quad $0.0506$ & \quad $0.0421\pm 0.00076$ \\ \hline
$\sin \theta^{(q)}_{13}$ & \quad $0.00354$ & \quad $0.00370$  & \quad $0.00370$  & \quad $0.00387$ & \quad $0.00365\pm 0.00012$ \\ \hline
$J_q$ & \quad $3.35\times 10^{-5}$ & \quad $3.50\times 10^{-5}$ & \quad $3.47\times 10^{-5}$ & \quad $2.97\times 10^{-5}$ & 
\quad $\left(3.18\pm 0.15\right)\times 10^{-5}$ \\ \hline
\end{tabular}%
\end{center}
\caption{Model benchmark scenarios S-4, S-3, S-2 with four, three and two free parameters, respectively, as well 
as experimental values of the quark sector observables from Ref.~\cite{Xing:2019vks,Tanabashi:2018oca}. 
We use the experimental values of the quark masses at the $M_{Z}$ scale from Ref.~\cite{Xing:2019vks}}.
\label{Tab}
\end{table}

Another observation about the set of values given in Eq. \eq{eq:bm-values} is that it shows rather particular pattern: some 
of them are practically equal between each other.  This fact suggests to consider the following simplified benchmark 
scenarios with a limited number of the free parameters:
\bea \label{eq:S-4}
\mbox{S-4 (4 free parameters):}&&\hspace{0.2cm}a_{11}^{(D) }=a_{12}^{(D) }=a_{22}^{(D)},\ \ \ a_1 ^{(U) }=a_{3}^{(U)  }=1,\ \ \
a_{23}^{(D)  }=a_{33}^{(D) }=a_{2}^{(U)  }.\\
\nn
\mbox{Best-fit values:}&&
\hspace{0.2cm}a_{2}^{(U)  }\simeq 1.40,\ \ \ a_{11}^{(D)
}\simeq 0.53, \ \ \ \left\vert a_{13}^{(D)  }\right\vert\simeq 0.43, \ \ \ \arg
(a_{13}^{( D) })\simeq -60.86^{\circ } \ \\
\label{eq:S-3}
\mbox{S-3 (3 free parameters):}&&\hspace{0.2cm}a_{11}^{(D)  }=a_{12}^{(D) }=a_{22}^{(D)  }=
\left\vert a_{13}^{(D)  }\right\vert,\ \ \ a_1 ^{(U) }=a_{3}^{(U)  }=1,\ \ \
a_{23}^{(D)  }=a_{33}^{(D)  }=a_{2}^{(U) }.\\
\nn
\mbox{Best-fit values:}&&
\hspace{0.2cm}a_{2}^{(U) }\simeq 1.40,\ \ \ a_{11}^{(D)
}\simeq 0.45, \ \ \ \arg
(a_{13}^{( D) })\simeq -60.9^{\circ }\ \\
\label{eq:S-2a}
\mbox{S-2a (2 free parameters):}&&\hspace{0.2cm}a_{11}^{(D)  }=a_{12}^{(D) }=a_{22}^{(D) }=
\left\vert a_{13}^{(D)  }\right\vert,\ \ \ a_1 ^{(U)  }=a_{3}^{(U) }=1,\ \ \
a_{23}^{(D)  }=a_{33}^{(D)  }=a_{2}^{(U)  }, \ \ \ \crn
\mbox{Best-fit values:}&&
\hspace{0.2cm}a_{2}^{(U) }\simeq 1.40,\ \ \ a_{11}^{(D)
}\simeq 0.45, \ \ \ \arg(a_{13}^{( D) })=-60^{\circ }.\\
\mbox{S-2b (2 free parameters):}&&\hspace{0.2cm}a_{11}^{(D)  }=a_{12}^{(D) }=a_{22}^{(D) }
=\left\vert a_{13}^{(D)  }\right\vert,\ \ \ a_1 ^{(U) }=a_{3}^{(U)  }=1,\ \ \
a_{23}^{(D)  }=a_{33}^{(D)  }=a_{2}^{(U)  }, \ \ \ \crn
\mbox{Best-fit values:}&&
\hspace{0.2cm}a_{2}^{(U)  }\simeq 1.40,\ \ \ a_{11}^{(D)
}\simeq 0.47, \ \ \ \arg(a_{13}^{( D) })=-45^{\circ }.
\nn
\eea
As seen from Table~\ref{Tab}, all the quark observables are reproduced with a reasonable precision even in the 2-parameter
 scenarios S-2a and S-2b. This result hints
that the model framework allows introduction of certain extra symmetries significantly reducing the number of free parameters. 
This possibility will be studied elsewhere.

\begin{figure}[]
		\centering
		\includegraphics[width=0.5\textwidth]{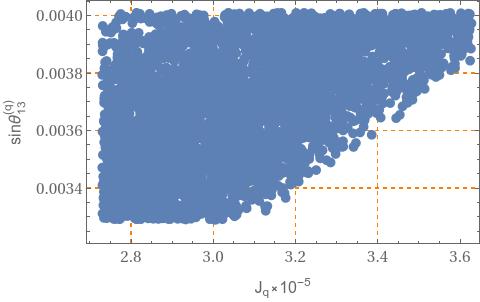}\newline
		\caption{Correlation of the quark mixing parameter $\sin\theta^{(q)}_{13}$ with the Jarlskog invariant.}
		\label{fig:correlationquarksector}
\end{figure}

Figure \ref{fig:correlationquarksector} shows the correlation of the quark mixing parameter $\sin\theta^{(q)}_{13}$ with the Jarlskog invariant. To obtain this figure, the quark sector parameters were randomly generated in a range of values where the CKM parameters and the quark masses are inside the $3\si $ experimentally allowed range. Such correlation shows that that the quark mixing parameter $\sin\theta^{(q)}_{13}$ and the Jarlskog invariant $J_q$ are located in the ranges $0.0033\lesssim\sin\theta^{(q)}_{13}\lesssim 0.0040$ and \mbox{$2.7\times 10^{-5}\lesssim J_q\lesssim 3.65\times 10^{-5}$}, respectively. We also found in this numerical analysis that the remaining quark mixing parameters are in the following ranges: $0.223\lesssim\sin\theta^{(q)}_{12}\lesssim 0.226$ and $0.040\lesssim\sin\theta^{(q)}_{23}\lesssim 0.045$.

Finally, the LHC signature of the exotic $T$, $J_{1}$ and $J_{2}$ quarks in our model is defined by the fact that they will mainly decay into a top quark plus neutral scalar and can be pair produced at the LHC via Drell-Yan and gluon fusion processes mediated by charged gauge bosons and gluons, respectively. 
Consequently, we consider the observation of an excess of events in the multijet and multilepton final state as the smoking gun of our model at the LHC. A detailed study of the collider phenomenology of the model is beyond the scope of this paper and is left for future studies.

	\section{Meson oscillations}
	
	\label{FCNC}It is worth mentioning that the non universal $U(1)_X$ charge assignments for the left handed quark fields give rise to flavour changing neutral processes (FCNC) mediated by the $Z^{\prime}$ gauge boson. These FCNC interactions contribute to the $K^0-\bar{K}^0$, $B^0_d-\bar{B}^0_d$ and $B^0_s-\bar{B}^0_s$ mass differences. It is worth mentioning that the $D^0-\bar{D}^0$ meson oscillations are absent at tree level since the symmetries of our model constrain the up type quark mass matrix to be diagonal. In this section we discuss the implications of our model in the
	Flavour Changing Neutral Current (FCNC) interactions in the down type quark
	sector. The flavour violating $Z^{\prime }$ interactions in the down type
	quark sector produce meson oscillations. The $K^{0}-\bar{K}^{0}$, $B_{d}^{0}-%
	\bar{B}_{d}^{0}$ and $B_{s}^{0}-\bar{B}_{s}^{0}$ meson mixings are described
	by the following effective Hamiltonians:%
	\be 
	\mathcal{H}_{eff}^{\left( K^{0}-\bar{K}^{0}\right) }\mathcal{=}\frac{4\sqrt{2%
		}G_{F}c_{W}^{4}m_{Z}^{2}}{\left( 3-4s_{W}^{2}\right) m_{Z^{\prime }}^{2}}%
	\left\vert \left( V_{DL}^{\ast }\right) _{32}\left( V_{DL}\right)
	_{31}\right\vert ^{2}O^{\left( K^{0}-\bar{K}^{0}\right) },
	\ee 
	\be 
	\mathcal{H}_{eff}^{\left( B_{d}^{0}-\bar{B}_{d}^{0}\right) }\mathcal{=}\frac{%
		4\sqrt{2}G_{F}c_{W}^{4}m_{Z}^{2}}{\left( 3-4s_{W}^{2}\right) m_{Z^{\prime
		}}^{2}}\left\vert \left( V_{DL}^{\ast }\right) _{31}\left( V_{DL}\right)
	_{33}\right\vert ^{2}O^{\left( B_{d}^{0}-\bar{B}_{d}^{0}\right) },
	\ee 
	\be 
	\mathcal{H}_{eff}^{\left( B_{s}^{0}-\bar{B}_{s}^{0}\right) }\mathcal{=}\frac{%
		4\sqrt{2}G_{F}c_{W}^{4}m_{Z}^{2}}{\left( 3-4s_{W}^{2}\right) m_{Z^{\prime
		}}^{2}}\left\vert \left( V_{DL}^{\ast }\right) _{32}\left( V_{DL}\right)
	_{33}\right\vert ^{2}O^{\left( B_{s}^{0}-\bar{B}_{s}^{0}\right) }.
	\ee 
	The $K^{0}-\bar{K}^{0}$, $B_{d}^{0}-\bar{B}_{d}^{0}$ and $B_{s}^{0}-\bar{B}%
	_{s}^{0}$ meson mixings in our model is caused by the tree level $Z^{\prime }
	$ exchange, thus giving generating the following operators: 
	\bea 
	O^{\left( K^{0}-\bar{K}^{0}\right) } &=&\left( \overline{s}\gamma _{\mu
	}P_{L}d\right) \left( \overline{s}\gamma ^{\mu }P_{L}d\right) ,\hspace{0.7cm}%
	\hspace{0.7cm}O^{\left( B_{d}^{0}-\bar{B}_{d}^{0}\right) }=\left( \overline{d%
	}\gamma _{\mu }P_{L}b\right) \left( \overline{d}\gamma ^{\mu }P_{L}b\right) ,
	\\
	O^{\left( B_{s}^{0}-\bar{B}_{s}^{0}\right) } &=&\left( \overline{s}\gamma
	_{\mu }P_{L}b\right) \left( \overline{s}\gamma ^{\mu }P_{L}b\right).
	\eea %
	Furthermore, the following relations have been taken into account: 
	\bea 
	\widetilde{M}_{f} &=&\left( M_{f}\right) _{diag}=V_{fL}^{\dagger
	}M_{f}V_{fR},\hspace{1cm}\hspace{1cm}f_{\left( L,R\right) }=V_{f\left(
		L,R\right) }\widetilde{f}_{\left( L,R\right) },  \crn 
	\overline{f}_{iL}\left( M_{f}\right) _{ij}f_{jR} &=&\overline{\widetilde{f}}%
	_{kL}\left( V_{fL}^{\dagger }\right) _{ki}\left( M_{f}\right) _{ij}\left(
	V_{fR}\right) _{jl}\widetilde{f}_{lR}=\overline{\widetilde{f}}_{kL}\left(
	V_{fL}^{\dagger }M_{f}V_{fR}\right) _{kl}\widetilde{f}_{lR}=\overline{%
		\widetilde{f}}_{kL}\left( \widetilde{M}_{f}\right) _{kl}\widetilde{f}%
	_{lR}=m_{fk}\overline{\widetilde{f}}_{kL}\widetilde{f}_{kR},  \crn
	k &=&1,2,3\,.
	\eea %
	Here, $\widetilde{f}_{k\left( L,R\right) }$ and $f_{k\left( L,R\right) }$ ($%
	k=1,2,3$) are the SM fermionic fields in the mass and interaction bases,	respectively.

It is worth mentioning as shown in detail in Appendix \ref{appHiggs}, that our model has the alignment limit for the lightest $126$ GeV SM like Higgs boson given that the remaining scalars are much heavier than the electroweak symmetry breaking scale $246$ GeV. Furthermore, our model at low energies, below the scale the scale of breaking of the $SU(3)_C\times SU(3)_L\times U(1)_X$ gauge symmetry, corresponds to a multiscalar singlet extension of the SM. Thus, the light $126$ GeV Higgs boson will not induce tree-level FCNC. 
This phenomenologically dangerous effect can happen in the presence of at least two SM doublet scalars before the electroweak symmetry breaking. To avoid this trouble, one can resort to  
the Glashow-Weinberg-Paschos theorem \cite{Glashow:1976nt,Paschos:1976ay} stating that there will be no tree-level FCNC coming from the scalar sector, if all right-handed fermions of a given electric charge couple to only one of the doublets.

	Besides that, the contributions to FCNC arising from the heavier scalars are strongly suppressed by their large mass scale and the very small mixings of the scalar singlets and the CP even neutral component of $\chi$ with the CP even electrically neutral component of $\eta$ (which is mostly composed of the $126$ GeV SM like Higgs boson). Because of this reason the FCNC interactions in our model mainly arise from the tree-level exchange of the $Z^{\prime }$ gauge boson. This situation is different than the one presented in 3-3-1 models with three scalar triplets like the ones considered in \cite{Huitu:2017ukq,Huitu:2019kbm,Huitu:2019mdr}, where two of the three scalar triplets do acquire VEVs at the electroweak symmetry breaking scale thus implying that at low energies below the TeV scale, the theory corresponds to a 2HDM where tree-level neutral scalar contributions to FCNC do exist. This problem was elegantly solved in Refs.~\cite{Huitu:2017ukq,Huitu:2019kbm,Huitu:2019mdr} by implementing the Froggatt-Nielsen mechanism in this version of the 3-3-1 model.
	
	On the other hand, the $K-\bar{K}$, $B_{d}^{0}-\bar{B}_{d}^{0}$ and $%
	B_{s}^{0}-\bar{B}_{s}^{0}$\ mass splittings are given by: 
	\be 
	\De m_{K}=\left( \Delta m_{K}\right) _{SM}+\Delta m_{K}^{\left( NP\right)
	},\hspace{1cm}\Delta m_{B_{d}}=\left( \Delta m_{B_{d}}\right) _{SM}+\De
	m_{B_{d}}^{\left( NP\right) },\hspace{1cm}\Delta m_{B_{s}}=\left( \De
	m_{B_{s}}\right) _{SM}+\De m_{B_{s}}^{\left( NP\right) },  \label{Deltam}
	\ee %
	where $\left( \De m_{K}\right) _{SM}$, $\left( \De m_{B_{d}}\right)
	_{SM}$ and $\left( \De m_{B_{s}}\right) _{SM}$ are the SM contributions,
	whereas $\De m_{K}^{\left( NP\right) }$ , $\De m_{B_{d}}^{\left(
		NP\right) }$ and $\left( \De m_{B_{s}}\right) _{SM}$ are new physics	contributions.
	
	In our model, the new physics contributions to the meson differences are
	given by: 
	\be 
	\De m_{K}^{\left( NP\right) }=\frac{4\sqrt{2}G_{F}c_{W}^{4}m_{Z}^{2}}{%
		\left( 3-4s_{W}^{2}\right) m_{Z^{\prime }}^{2}}\left\vert \left(
	V_{DL}^{\ast }\right) _{32}\left( V_{DL}\right) _{31}\right\vert
	^{2}f_{K}^{2}B_{K}\eta _{K}m_{K},
	\ee %
	\be 
	\De m_{B_{d}}^{\left( NP\right) }=\frac{4\sqrt{2}G_{F}c_{W}^{4}m_{Z}^{2}}{%
		\left( 3-4s_{W}^{2}\right) m_{Z^{\prime }}^{2}}\left\vert \left(
	V_{DL}^{\ast }\right) _{31}\left( V_{DL}\right) _{33}\right\vert
	^{2}f_{B_{d}}^{2}B_{B_{d}}\eta _{B_{d}}m_{B_{d}},
	\ee %
	\be 
	\Delta m_{B_{s}}^{\left( NP\right) }=\frac{4\sqrt{2}G_{F}c_{W}^{4}m_{Z}^{2}}{%
		\left( 3-4s_{W}^{2}\right) m_{Z^{\prime }}^{2}}\left\vert \left(
	V_{DL}^{\ast }\right) _{32}\left( V_{DL}\right) _{33}\right\vert
	^{2}f_{B_{s}}^{2}B_{B_{s}}\eta _{B_{s}}m_{B_{s}}.
	\ee %
	Using the following parameters \cite{Dedes:2002er,Aranda:2012bv,Khalil:2013ixa,Queiroz:2016gif,Buras:2016dxz,Ferreira:2017tvy,Duy:2020hhk}: 
	\bea 
	\Delta m_{K} &=&\left( 3.484\pm 0.006\right) \times 10^{-12}MeV,\hspace{1.5cm%
	}\left( \Delta m_{K}\right) _{SM}=3.483\times 10^{-12}MeV  \crn 
	f_{K} &=&160MeV,\hspace{1.5cm}B_{K}=0.85,\hspace{1.5cm}\eta _{K}=0.57,%
	\hspace{1.5cm}m_{K}=497.614MeV.  \nn 
	\eea%
	\bea
	\left( \Delta m_{B_{d}}\right) _{\exp } &=&\left( 3.337\pm 0.033\right)
	\times 10^{-10}MeV,\hspace{1.5cm}\left( \Delta m_{B_{d}}\right)
	_{SM}=3.582\times 10^{-10}MeV,  \crn 
	f_{B_{d}} &=&188MeV,\hspace{1.5cm}B_{B_{d}}=1.26,\hspace{1.5cm}\eta
	_{B_{d}}=0.55,\hspace{1.5cm}m_{B_{d}}=5279.5MeV.  \nn
	\eea %
	\bea 
	\left( \Delta m_{B_{s}}\right) _{\exp } &=&\left( 104.19\pm 0.8\right)
	\times 10^{-10}MeV,\hspace{1.5cm}\left( \Delta m_{B_{s}}\right)
	_{SM}=121.103\times 10^{-10}MeV,  \crn 
	f_{B_{s}} &=&225MeV,\hspace{1.5cm}B_{B_{s}}=1.26,\hspace{1.5cm}\eta
	_{B_{s}}=0.55,\hspace{1.5cm}m_{B_{s}}=5366.3MeV.  \nn
	\eea%
	\begin{figure}[h]
		\includegraphics[width=0.51\textwidth]{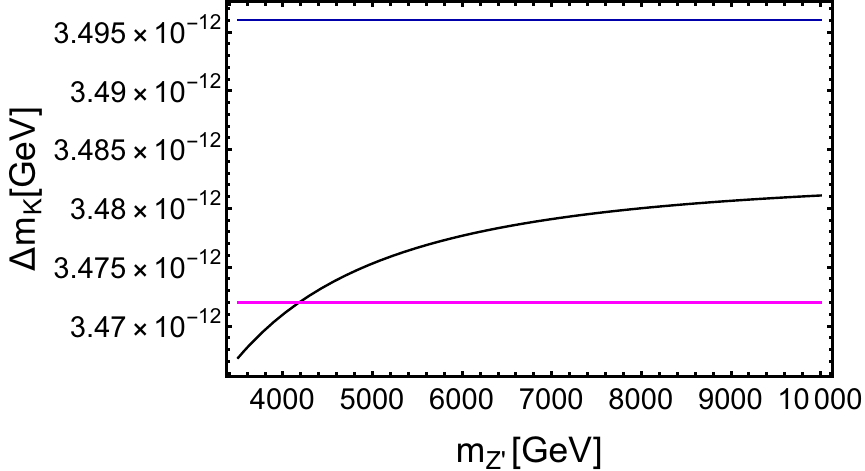}\includegraphics[width=0.51%
		\textwidth]{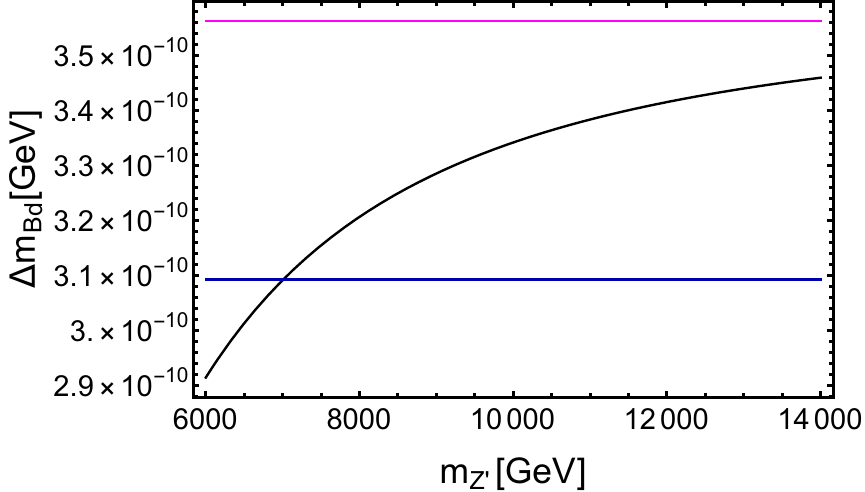}\newline
		\includegraphics[width=0.51\textwidth]{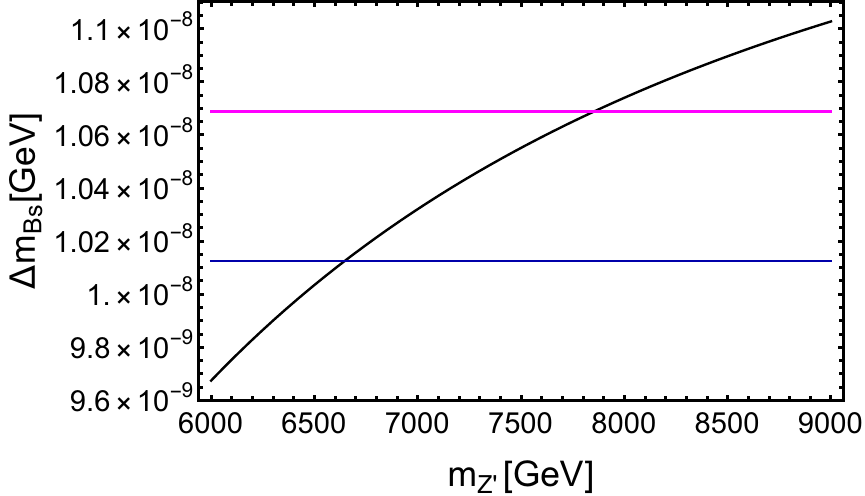}
		\caption{The $K^{0}-\bar{K}^{0}$, $B_{d}^{0}-\bar{B}_{d}^{0}$ and $B_{s}^{0}-\bar{B}_{s}^{0}$\ mass splittings as as function of the $Z^{\prime }$ mass.}
		\label{mesons}
	\end{figure}
	We plot in figure \ref{mesons} the $K^{0}-\bar{K}^{0}$, $B_{d}^{0}-\bar{B}%
	_{d}^{0}$ and $B_{s}^{0}-\bar{B}_{s}^{0}$\ mass splittings as as function of
	the $Z^{\prime }$ mass. As seen from figure \ref{mesons}, the $K^{0}-\bar{K}%
	^{0}$, $B_{d}^{0}-\bar{B}_{d}^{0}$ and $B_{s}^{0}-\bar{B}_{s}^{0}$
	oscillations caused by the flavor changing neutral interactions reach values
	close to their experimental upper limits and the constraints arising from
	these meson oscillations set the $Z^{\prime }$ mass in the range $7$ TeV$%
	\lesssim m_{Z^{\prime }}\lesssim 8$ TeV.

\newpage

\section{Lepton masses and mixings}
\label{leptonsector}
From the charged lepton Yukawa terms, we find the charged lepton mass matrix in the form:
\bea
M_l &=&\resizebox{0.95\hsize}{!}{\fontsize{3}{3}{{%
\tiny$\left(
\begin{array}{ccc}
 a_1\la ^9+b_1\la ^9 \left(\cos (\al )-e^{-i \psi } \sin (\al )\right) & b_2\la ^5 \left(\cos
   (\al )+e^{i \psi } \sin (\al )\right) & a_3\la ^3+b_3\la ^3 \left(\cos (\al )-e^{-i \psi } \sin (\al )\right)
   \\
 a_1\la ^9+b_1\la ^9 \om ^2 \left(\cos (\al )-e^{-i \psi } \om ^2 \sin (\al )\right) & b_2\la ^5
 \om \left(\cos (\al )+e^{i \psi } \om \sin (\al )\right) & a_3\la ^3+b_3\la ^3 \om ^2 \left(\cos
   (\al )-e^{-i \psi } \om ^2 \sin (\al )\right) \\
 a_1\la ^9+b_1\la ^9 \om \left(\cos (\al )-e^{-i \psi } \om \sin (\al )\right) & b_2\la ^9
 \om ^2 \left(\cos (\al )+e^{i \psi } \om ^2 \sin (\al )\right) & a_3\la ^3+b_3\la ^3 \om \left(\cos
   (\al )-e^{-i \psi } \om \sin (\al )\right) \\
\end{array}
\right)$}}}\crn 
&&\times\fr{v_{\eta}}{\sqrt{2}},
\label{leptonmasses}
\eea%
where $a_1$, $a_3$, $b_i$ ($i=1,2,3$) are $\mathcal{O}(1)$
parameters constructed of the parameters $y^{(L)}_{i}, z^{(L)}$. Note that the charged lepton masses are linked to the
scale of the electroweak symmetry breaking through their power dependence on the
Wolfenstein parameter $\la $, with $\mathcal{O}(1)$ coefficients.
Furthermore, from the lepton Yukawa terms given in Eq. \eq{Lyl} it
follows that our model does not feature flavor changing leptonic neutral
Higgs decays at tree level.

For the neutrino sector we find from Eq. \eq{Lyl} the
neutrino mass term:
\be
- 2\mathcal{L}_{mass}^{( \nu ) }=\left(
\begin{array}{ccc}
\overline{\nu _L^{\ C}} & \overline{\nu _R } & \overline{N_R }%
\end{array}%
\right) M_\nu \left(
\begin{array}{c}
\nu_L \\
\nu_R^{\ C} \\
N_R^{\ C}%
\end{array}%
\right) +H.c,  \label{Lnu}
\ee%
where $\nu_{iR}\equiv ((\nu^c)_L)^C$ corresponds to the third components of the lepton triplet introduced in Eq.~\eq{L}. 
The $A_4$ family symmetry of the model constrains the neutrino mass matrix to be
of the form:
\be
M_\nu =\left(
\begin{array}{ccc}
0_{3\times 3} & M_1  & M_2 \\
M_1 ^T  & 0_{3\times 3} & M_3  \\
M_2^T  & M_3^T  & 0_{3\times 3}%
\end{array}%
\right)  \label{Mnu0}
\ee
with
\bea  \label{M123}
M_1  &=&\fr{v_\eta v_\chi v_\zeta }{2\sqrt{2}\La^2}\left(
\fr{v_\si }{\La }\right) ^{11}\left(
\begin{array}{ccc}
0 & \om^2 & 0 \\
-\om^2 & 0 & 1 \\
0 & -1 & 0%
\end{array}%
\right) ,\cm \cm  M_2=y_{1\eta }^{( L) }\fr{
v_\eta v_\xi  }{\sqrt{6}\La }\left( \fr{ v_\si } \La %
\right)\left(
\begin{array}{ccc}
0 & ( 1+x) \om^2 & ( 1-x) \om \\
( 1-x) \om^2 & 0 & 1+x \\
( 1+x) \om & 1-x & 0%
\end{array}
\right),  \crn
M_3  &=&y_\chi^{( L) }\fr{v_\chi }{\sqrt{2}}\left( \fr{%
v_\si }{\La }\right)\left(
\begin{array}{ccc}
1 & 0 & 0 \\
0 & \om & 0 \\
0 & 0 & \om^2%
\end{array}%
\right) ,\cm \cm x=\fr{y_{2\eta }^{( L) }}{%
y_{1\eta }^{( L) }},\cm \cm \om =e^{\fr{2\pi
i}{3}}.
\eea

The light active masses arise from linear seesaw mechanism and the
physical neutrino mass matrices are:

\bea
M_\nu ^{( 1) } &=&-\left[ M_2M_3 ^{-1}M_1 ^T +M_1 \left(
M_3^T \right) ^{-1}M_2^T \right] \,,  \label{M1nu} \\
M_\nu ^{( 2) } &=&-\fr 1 2 \left( M_3 +M_3^T \right) +%
\fr{1}{4}\left[ M_1^T(M_3^\ast )^{-1}M_2^{\ast }+M_2^{\dagger}(M_3^{\dagger} )^{-1} M_1\right] \,, \label{M2nu}\\
M_\nu^{( 3) } &=&\fr 1 2 \left( M_3 +M_3 ^T \right)+\fr{1}{4}\left[ M_1^T(M_3^\ast )^{-1}M_2^{\ast }+M_2^{\dagger}(M_3^{\dagger} )^{-1} M_1\right] \,%
 \,, \label{M3nu}
\eea%
where $M_\nu ^{( 1) }$ is the active neutrino mass
matrix whereas $M_\nu ^{( 2) }$ and $M_\nu^{( 3)} $ are the sterile neutrino mass matrices.
Explicitly we have

\bea  \label{Mnu}
M_\nu^{( 1) } &=&\fr{y_{1\eta }^{( L) }}{\sqrt{2}%
y_\chi ^{( L) }}\left( \fr{v_\si }{\La }\right) ^{11}%
\fr{v_\eta v_\zeta v_\xi  }{\La ^3 }\left(
\begin{array}{ccc}
-2(x+1) & \om ^2(x-1) & 2\om x \\
\om ^2(x-1) & -4\om x & x+1 \\
2\om x & x+1 & -2\om ^2(x-1)
\end{array}%
\right) \fr{v_\eta }{\sqrt{2}}  \crn
&=&\left(
\begin{array}{ccc}
-2(x+1) & \om^2(x-1) & 2\om x \\
\om ^2(x-1) & -4\om x & x+1 \\
2\om x & x+1 & -2\om ^2(x-1)
\end{array}%
\right) m_\nu ,\cm m_\nu =\fr{a_\nu \la ^{19}v}{\sqrt{2}}\,.
\eea
The experimental values of charged lepton masses, the neutrino mass squared splittings, the
leptonic mixing parameters  and Dirac CP violating phase can be reproduced
for the normal ordering (NO) of the neutrino mass spectrum with the following values of the model effective parameters:
\be
\begin{array}{c}
a_1\simeq 0.983,\cm a_3\simeq -0.483,\cm b_1\simeq -0.755,\cm b_2\simeq -0.597,\cm b_3\simeq -0.199,\cm x\simeq 0.431,\\
m_\nu \simeq 16.34\ \mbox{meV},\cm \al
\simeq 122.25^{\circ }\,,\cm \beta \simeq -42.82^{\circ }\,,\cm%
\ga \simeq -59.36^{\circ }\,,\cm%
\psi \simeq 98.44^{\circ }\,.%
\end{array}
\label{Benchmarklepton}
\ee
Using the values of the lepton model effective parameters of Eq. (\ref{Benchmarklepton}), the PMNS leptonic mixing matrix takes the form:
\be
U_{PMNS}=U^{\+ }_{l}U_{\nu}=\left(
\begin{array}{ccc}
 -0.818231-0.0686404 i & -0.318382+0.449127 i & 0.148954\, +0.0227392 i \\
 0.0515003\, +0.373766 i & -0.379958-0.371145 i & 0.731222\, +0.202101 i \\
 -0.180118-0.388575 i & 0.634519\, -0.110392 i & 0.564605\, -0.288074 i \\
\end{array}
\right)
\label{U}
\ee
where:
\bea
U_l&=&\left(
\begin{array}{ccc}
 -0.625827& -0.614417 & -0.48045 \\
 -0.406057-0.298325 i & 0.726662\, +0.202306 i & -0.400359+0.129878 i \\
 0.569876\, -0.172341 i & -0.112245-0.202306 i & -0.59877+0.483205 i \\
\end{array}
\right),\crn
U_{\nu}&=&\left(
\begin{array}{ccc}
 0.566967 & 0.12785 & -0.813759 \\
 0.396158\, +0.686167 i & -0.177447-0.307348 i & 0.248135\, +0.429782 i \\
 -0.112675+0.195159 i & -0.463062+0.802046 i & -0.151256+0.261983 i \\
\end{array}
\right).
\label{UlUnu}
  \eea
As seen from Table~\ref{Neutrinos}, the model values are consistent with the experimental ones. 
Again, akin to the quark sector, the absolute value of the effective dimensionless parameters $a^{(l)}, x$ are of the order of unity. 
We interpret this fact in a way that the lepton mass hierarchy is explained on account of  the model structure, symmetries and field
 content, without unnatural tuning these effective parameters.
\begin{table}[th]
\begin{center}
\renewcommand{\arraystretch}{1}
\begin{tabular}{c|l|l|l|l|l|l|}
\hline
Observable & Model & bpf $\pm 1\si$ \cite{deSalas:2017kay} & bpf $\pm
1\si$ \cite{Esteban:2016qun} & $2\si$ range \cite{deSalas:2017kay} & $3\si$ range \cite{deSalas:2017kay} & $%
3\si$ range \cite{Esteban:2016qun} \\ \hline
$\De m_{21}^2$ [$10^{-5}$eV$^2$] & \quad $7.59$ & \quad $%
7.55_{-0.16}^{+0.20}$ & \quad $7.40_{-0.20}^{+0.21}$ & \quad $7.20-7.94$ & \quad $7.05-8.14$ &
\quad $6.80-8.02$ \\ \hline
$\De m_{31}^2$ [$10^{-3}$eV$^2$] & \quad $2.53$ & \quad $2.50\pm 0.03$
& \quad $2.494_{-0.031}^{+0.033}$ & \quad $2.44-2.57$ & \quad $2.41-2.60$ & \quad $2.399-2.593$
\\ \hline
$\theta^{(l)}_{12} (^{\circ })$ & \quad $33.84$ & \quad $34.5_{-1.0}^{+1.2}$
& \quad $36.62_{-0.76}^{+0.78}$ & \quad $32.5-36.8$ & \quad $31.5-38.0$ & \quad $31.42-36.05$ \\
\hline
$\theta^{(l)}_{13} (^{\circ })$ & \quad $8.67$ & \quad $8.45_{-0.14}^{+0.16}$
& \quad $8.54\pm 0.15$ & \quad $8.2-8.8$ & \quad $8.0-8.9$ & \quad $8.09-8.98$ \\ \hline
$\theta^{(l)}_{23} (^{\circ })$ & \quad $50.12$ & \quad $47.9_{-1.7}^{+1.0}$
& \quad $47.2_{-3.9}^{+1.9}$ & \quad $43.1-49.8$ & \quad $41.8-50.7$ & \quad $40.3-51.5$ \\
\hline
$\de^{(l)}_{CP} (^{\circ })$ & \quad $-85.29$ & \quad $-142_{-27}^{+38}$ &
\quad $-108_{-31}^{+43}$ & \quad $182-315$ & \quad $157-349$ & \quad $144-374$ \\ \hline\hline
\end{tabular}%
\end{center}
\caption{Model and experimental values of the light active neutrino masses,
leptonic mixing angles and CP violating phase for the scenario of normal
(NH) neutrino mass hierarchy. The experimental values are taken from Refs.
\protect\cite{deSalas:2017kay,Esteban:2016qun}}
\label{Neutrinos}
\end{table}

\begin{figure}[]
	\centering
	\includegraphics[width=0.5\textwidth]{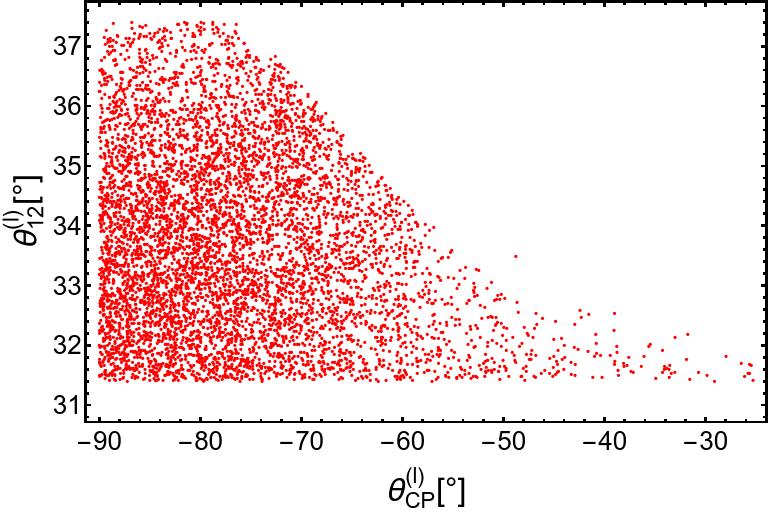}\includegraphics[width=0.5\textwidth]{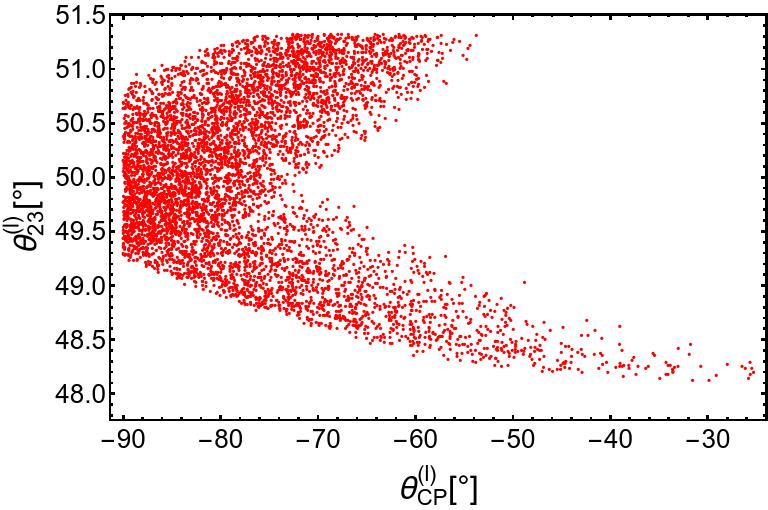}\newline
	\includegraphics[width=0.5\textwidth]{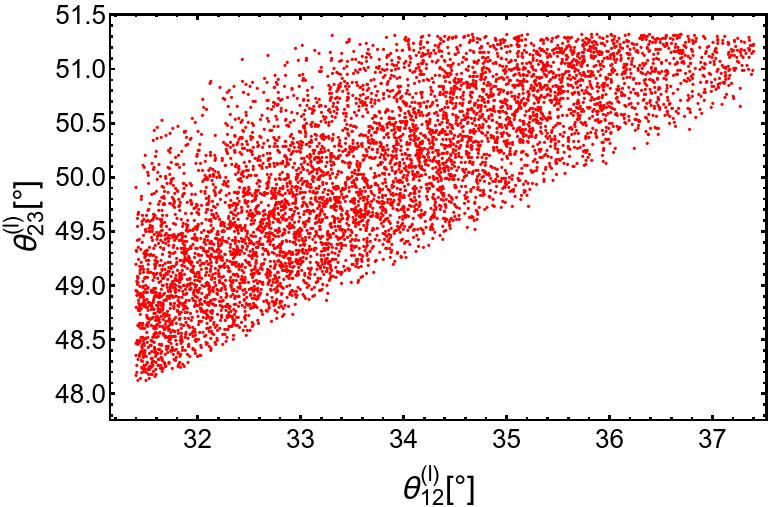}\includegraphics[width=0.5\textwidth]{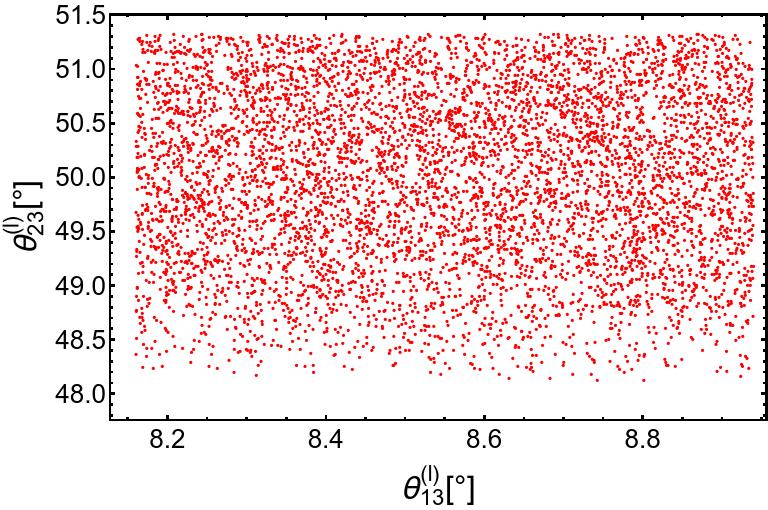}\newline
	\caption{Correlations between the different lepton sector observables.}
	\label{fig:correlationleptonsector}
\end{figure}

Figure \ref{fig:correlationleptonsector} shows the correlations of the leptonic mixing angles with the leptonic Dirac CP-violating phase as
 well as the correlations between the leptonic mixing parameters. To obtain these Figures, the lepton sector parameters were randomly 
 generated in a range of values where the neutrino mass squared splittings, leptonic mixing parameters and leptonic Dirac CP violating
  phase are consistent with the experimental data. These lepton sector observables are inside the $1\si $ experimentally allowed range,
   excepting $\theta^{(l)}_{23}$ which is inside the $3\si $ range. We found the leptonic Dirac CP violating phase in the 
   range $-90^{\circ}\lesssim\delta^{(l)}_{CP}\lesssim -25^{\circ}$, whereas the leptonic mixing angles are obtained to be in the ranges $31.5^{\circ}\lesssim\theta^{(l)}_{12}\lesssim 37.5^{\circ}$, $48.0^{\circ}\lesssim\theta^{(l)}_{23}\lesssim 51.5^{\circ}$ and $8.15^{\circ}\lesssim\theta^{(l)}_{12}\lesssim 8.9^{\circ}$.

Let us consider the effective Majorana neutrino mass parameter
\be
m_{\beta\beta}=\left\vert \sum_j U_{ek}^2m_{\nu _k}\right\vert ,  \label{mee}
\ee%
where $U_{ej}$ and $m_{\nu _k}$ are the the PMNS leptonic mixing matrix
elements and the neutrino Majorana masses, respectively. The neutrinoless
double beta ($0\nu \beta \beta $) decay amplitude is proportional to $m_{\beta\beta}$.

\begin{figure}[]
	\centering
	\includegraphics[width=0.5\textwidth]{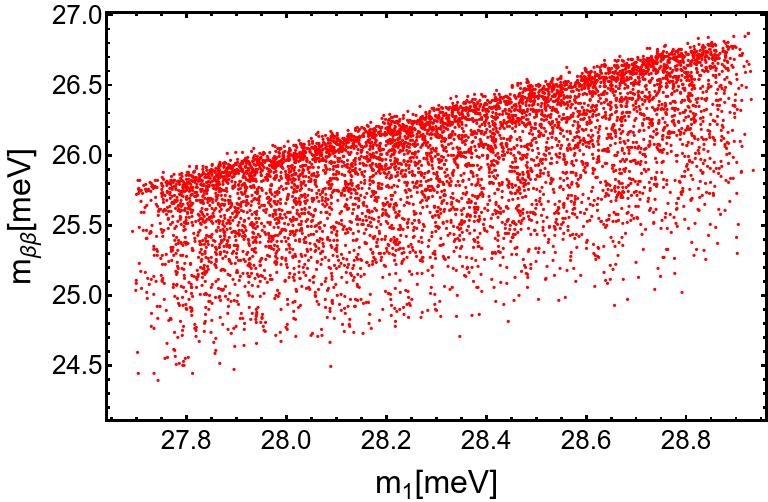}\includegraphics[width=0.5\textwidth]{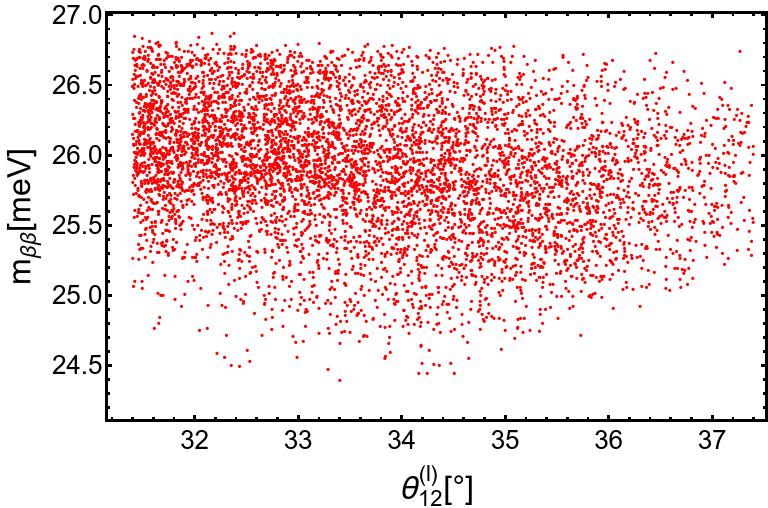}\newline
   \includegraphics[width=0.5\textwidth]{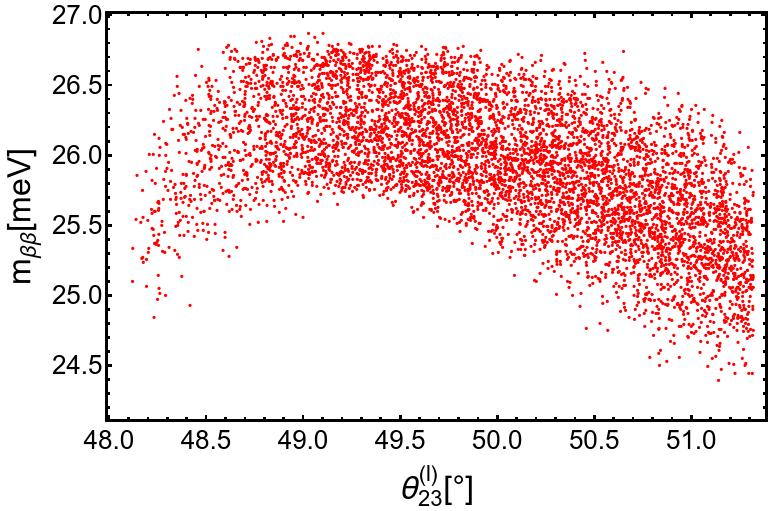}\includegraphics[width=0.5\textwidth]{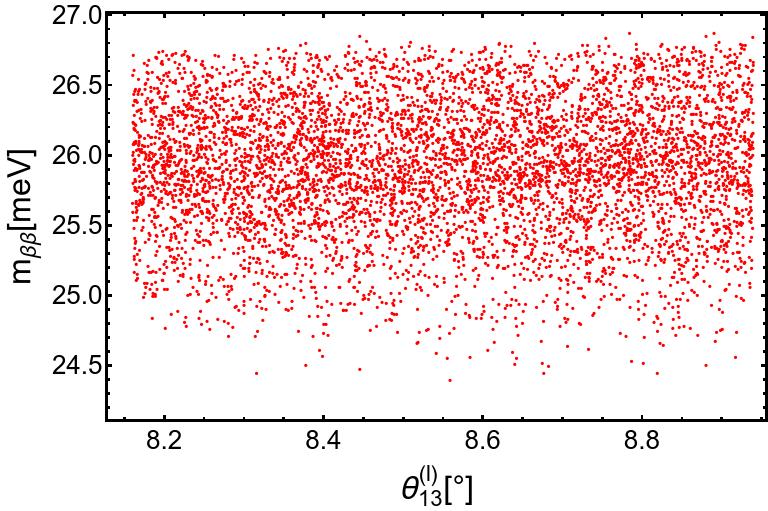}\newline	
	\caption{Correlations of the effective Majorana neutrino mass parameter	$m_{\beta\beta}$ with the lightest neutrino mass $m_1$ and with the leptonic mixing parameters.}
	\label{Correlationmee}
\end{figure}

Fig.~\ref{Correlationmee} shows the correlation of the effective Majorana neutrino mass parameter $m_{ee}$ vs the lightest neutrino mass $m_1$.
As can be seen from Fig.~\ref{Correlationmee}, our model
predicts the values of the effective Majorana neutrino mass parameter in the range $24.5$~meV$\lesssim m_{\beta\beta}\lesssim$ $27$ meV,
which is within the declared reach of the next-generation bolometric CUORE
experiment \cite{Alessandria:2011rc} or, more realistically, of the
next-to-next-generation ton-scale $0\nu \beta \beta $-decay experiments. The
current most stringent experimental upper limit
$m_{\beta\beta}\leq 160$ meV
is set
by $T_{1/2}^{0\nu \beta \beta }(^{136}\mathrm{Xe})\geq 1.1\times 10^{26}$ yr
at 90\% C.L. from the KamLAND-Zen experiment \cite{KamLAND-Zen:2016pfg}.

\section{$Z^\prime$ gauge boson production at the LHC}
\label{zlhc}
Here we compute the total cross section
for the production of the heavy $Z^\prime $ gauge boson, defined in Eq~(\ref{eq:Neutral-Bosons}), 
at the LHC via Drell-Yan mechanism. We consider the dominant contribution due to the parton distribution
functions of the light up, down and strange quarks, so that the total cross
section for the production of a $Z^\prime $ via quark antiquark
annihilation in proton-proton collisions with center of mass energy $\sqrt{S}
$ takes the form:
\small{
\bea
\si _{pp\rightarrow Z^\prime }^{\left( DrellYan\right) }( S)
&=&\fr{g^2 \pi }{6c_W^2 S}\left\{ \left[ \left( g_{uL}^\prime
\right) ^2 +\left( g_{uR}^\prime \right) ^2 \right] \int_{\ln \sqrt{%
\fr{m_{Z^\prime }^2 }{S}}}^{-\ln \sqrt{\fr{m_{Z^\prime }^2 }{S}}%
}f_{p/u}\left( \sqrt{\fr{m_{Z^\prime }^2 }{S}}e^y,\mu^2 \right)
f_{p/\overline{u}}\left( \sqrt{\fr{m_{Z^\prime }^2 }{S}}e^{-y},\mu
^2 \right) dy\right.  \crn
&&+\left. \left[ \left( g_{dL}^\prime \right) ^2 +\left( g_{dR}^\prime
\right)^2 \right] \int_{\ln \sqrt{\fr{m_{Z^\prime }^2 }{S}}}^{-\ln
\sqrt{\fr{m_{Z^\prime }^2 }{S}}}f_{p/d}\left( \sqrt{\fr{m_{Z^\prime
}^2 }{S}}e^{y},\mu ^2 \right) f_{p/\overline{d}}\left( \sqrt{\fr{%
m_{Z^\prime }^2 }{S}}e^{-y},\mu ^2 \right) dy\right.  \crn
&&+\left. \left[ \left( g_{dL}^\prime \right) ^2 +\left( g_{dR}^\prime
\right) ^2 \right] \int_{\ln \sqrt{\fr{m_{Z^\prime }^2 }{S}}}^{-\ln
\sqrt{\fr{m_{Z^\prime }^2 }{S}}}f_{p/s}\left( \sqrt{\fr{m_{Z^\prime
}^2 }{S}}e^y,\mu^2 \right) f_{p/\overline{s}}\left( \sqrt{\fr{%
m_{Z^\prime }^2 }{S}}e^{-y},\mu^2 \right) dy\right\}\, ,
\eea%
where $g_{uL(R)}^\prime$, $g_{dL(R)}^\prime$ are the $Z^\prime $ couplings to left (right) handed up and down type quarks, respectively.
These couplings are given in Appendix \ref{app_Zcoupling}.
The functions $f_{p/u}\left( x_1,\mu^2 \right)$ ($f_{p/\overline{u}}\left(x_2,\mu ^2 \right)$), $f_{p/d}\left(x_1,\mu
^2 \right)$ ($f_{p/\overline{d}}\left( x_2,\mu^2 \right)$) and $f_{p/s}\left(x_1,\mu
^2 \right)$ ($f_{p/\overline{s}}\left( x_2,\mu^2 \right)$) are the distributions of the light up, down and strange quarks (antiquarks), respectively,
 in the proton which carry
momentum fractions $x_1$ ($x_2$) of the proton.

The factorization scale is taken to be $\mu =m_{Z^\prime }$.
\begin{figure}[tbh]
\resizebox{8.7cm}{9cm}{\includegraphics{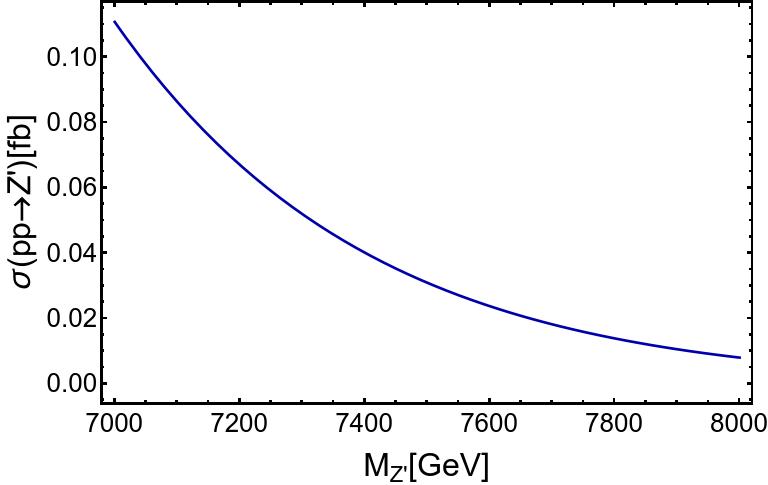}}\resizebox{8.7cm}{9cm}{\includegraphics{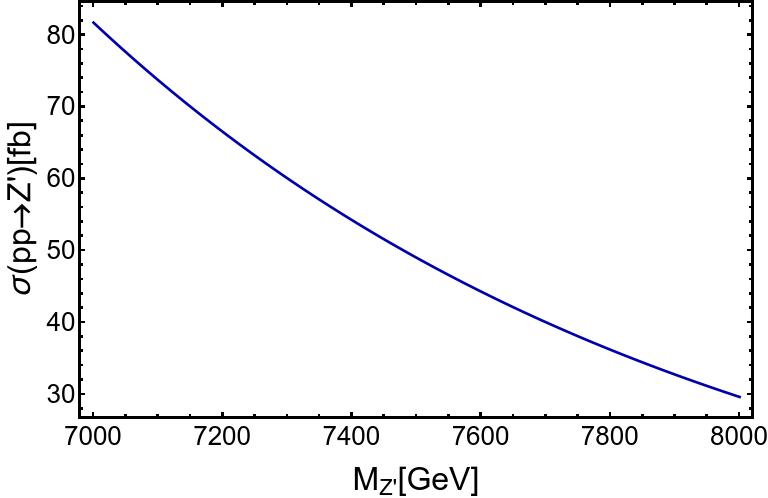}}
\caption{Total cross section for the $Z^\prime $ production via Drell-Yan
		mechanism at the LHC for $\protect\sqrt{S}=13$ TeV (left plot) and $\protect\sqrt{S}=28$ TeV (right plot) and as a function of the $%
		Z^\prime $ mass.}
\label{qqtoZprime}
\end{figure}

Fig.~\ref{qqtoZprime} (left panel) displays the $Z^\prime $ total production cross section at the
LHC via the Drell-Yan mechanism for $\protect\sqrt{S}=13$ TeV as a function of the $Z^\prime $ mass $M_{Z'}$
in the range from $7$ TeV up to $8$ TeV. We consider $M_{Z'}\geq 7$ TeV in order to fulfill
the bound arising from the experimental data on $K$, $B_d$ and $B_s$ meson mixings obtained in section \ref{FCNC}
For this region of $Z^\prime $ masses we find that the total production cross section ranges from $0.11$ fb up to $0.01$ fb.
The heavy neutral $Z^\prime$ gauge boson, after being produced, will subsequently decay into the pair of the SM particles,
with the dominant decay mode into quark-antiquark pairs as shown in Refs. \cite{Perez:2004jc,CarcamoHernandez:2005ka}. 
The two body decays of the $Z^\prime$ gauge boson in 3-3-1 models have been studied in details in Ref. \cite{Perez:2004jc}.
In particular, in Ref. \cite{Perez:2004jc} it has been shown that in 3-3-1 models the $Z^\prime$ decays into a lepton pair 
have branching ratios of the order of $10^{-2}$, which implies that the total LHC cross section for the $pp\to Z^\prime\to l^+l^-$ 
resonant production at $\protect\sqrt{S}=13$ TeV will be of the order of $10^{-3}$ fb for a $7$ TeV $Z^\prime$ gauge boson,
which is below its corresponding lower experimental limit from the LHC searches \cite{Aaboud:2017sjh}.
On the other hand, at the proposed energy upgrade of the LHC up to 28 TeV center of mass energy, the total cross section 
for the Drell-Yan production of a heavy $Z^\prime $ neutral gauge boson gets significantly enhanced reaching values
ranging from $82$ fb up to $30$ fb, as indicated in the right panel of Fig.~\ref{qqtoZprime}. Consequently, the LHC cross section for the $pp\to Z^\prime\to l^{+}l^{-}$ resonant production at $\protect\sqrt{S}=28$ TeV will be of the order of $1$ fb for a $7$ TeV $Z^\prime$ gauge boson, which is consisteny with its corresponding lower experimental limit arising from the LHC searches \cite{Aaboud:2017sjh}.

\section{Lepton flavour violating decays}
\label{lfvdecay}

Let us analyze the implications of our model for the LFV decays of the SM charged leptons and Higgs boson.

Given that the SM charged lepton mass matrix (\ref{leptonmasses})
cannot be diagonalized analytically in the practically useful form, in this section, for the sake of simplicity, 
we restrict ourselves to a simplified benchmark scenario characterized by the relations:

\bea \label{eq:benchmark-1}
&& z_1^{( L) }=y_1 ^{( L) },\ \  v_\phi
=e^{i\ga }\sin \beta v_{\va }, \ \  v_\rho =v_\va \cos
\beta ,\ \  y_3 ^{( L) }=-e^{-i\ga }y_4^{(
L) }\tan \beta ,\cm z_3^{( L) }=e^{-i\ga
}y_4^{( L) }\cot \beta .
\eea
Then, the charged lepton mass matrix takes the form:
\bea
M_l &=&R_{lL}diag\left( m_e, m_\mu , m_\tau \right) ,\cm R_{lL}=%
\fr{1} {\sqrt{3}}\left(
\begin{array}{ccc}
1 & 1 & 1 \\
1 & \om & \om^2 \\
1 & \om^2 & \om%
\end{array}%
\right) \left(
\begin{array}{ccc}
1 & 0 & 0 \\
0 & \cos \al & -\sin \al e^{-i\psi }\\
0 & \sin \al e^{i\psi } & \cos \al%
\end{array}%
\right)
\left(
\begin{array}{ccc}
\cos \beta & 0 & -\sin \beta e^{-i\ga } \\
0 & 1 & 0 \\
\sin \beta e^{i\ga } & 0 & \cos\beta%
\end{array}%
\right),  \crn
\om &=&e^{\fr{2\pi i}{3}},  \label{Ml}
\eea%
where the charged lepton masses are:
\be
m_e =a_1 ^{( l) }\la ^{9}\fr{v}{\sqrt{2}},\cm %
m_\mu =a_2^{( l) }\la ^{5}\fr{v}{\sqrt{2}},\cm %
m_\tau =a_3 ^{( l) }\la ^3 \fr{v}{\sqrt{2}}.
\label{leptonmasses}
\ee

In Appendix \ref{appHiggs} we derived an expression (\ref{h012}) for the SM Higgs boson, $h^{0}_{1}$, as 
a linear combination of the scalars present in our model.
We combine such relations with the definitions of the charged lepton mass eigenstates and masses:
\bea
\widetilde{M}_{f} &=&\left( M_{f}\right) _{diag}=V_{fL}^{\+ }M_{f}V_{fR},%
\hspace{1cm}\hspace{1cm}f_{\left( L,R\right) }=V_{f\left(L,R\right) }
\widetilde{f}_{\left( L,R\right) },  \notag \\
\overline{f}_{iL}\left( M_{f}\right) _{ij}f_{jR} &=&\overline{\widetilde{f}}%
_{kL}\left( V_{fL}^{\+  }\right) _{ki}\left( M_{f}\right)_{ij}
\left(V_{fR}\right) _{jl}\widetilde{f}_{lR}=\overline{\widetilde{f}}%
_{kL}\left(V_{fL}^{\+  }M_{f}V_{fR}\right) _{kl}\widetilde{f}_{lR}=
\overline{\widetilde{f}}_{kL}\left( \widetilde{M}_{f}\right) _{kl}\widetilde{%
f}_{lR}=m_{fk}\overline{\widetilde{f}}_{kL}\widetilde{f}_{kR},  \notag \\
k &=&1,2,3 \,.
\eea
where $\widetilde{f}_{k\left( L,R\right) }$ and $f_{k\left( L,R\right) }$ ($%
k=1,2,3$) are the SM fermion mass and interaction eigenstates, respectively.

Then, considering the first three terms in Eq.~ (\ref{Lyl}) we find the $h^0_1ee$ couplings
\begin{align}  \label{h01ee}
-\mathcal{L}_{h^0_1 ee}&\subset \left( 1+ \fr{\xi_\eta }{v_\eta } +
\fr{\xi_\chi }{v_\chi }\right) \left( m_{e_i} \overline{e_{iL}} e_{iR} +%
\mathrm{H.c.}\right) \rightarrow \fr{g}{2 m_W} \left( c_\al
+s_{\al} t_\theta\right) h^0_1 \left( m_{e_i} \overline{e_{iL}} e_{iR} +%
\mathrm{H.c.}\right) ,
\end{align}
coinciding in the limit $s_{\al} \to 0$ with the SM ones.
As seen from the above formula, there are no lepton flavor violating decays
of the SM-like Higgs bosons (LFVHD) $h^0_1\rightarrow e_i^\pm e_j^{\mp}$
with $i\neq j$ at tree level.
This is consistent with the latest
experimental result, where no signals were found setting the upper bound Br$(h^0_1\rightarrow \tau^{\mp}\mu^{\pm},\tau^{\mp} e^{\pm}) <%
\mathrm{O}(10^{-3})$ at 95 \% confidence level~\cite{Sirunyan:2017xzt,Aad:2019ugc}. This feature distinguishes our model
from some previous models with discrete symmetry that predicted tree-level LFVHD~\cite{Campos:2014zaa}.
 However, the SM-like Higgs bosons in our model still couple with the heavy neutrinos through the four last Yukawa 
 terms in Eq.~\eq{Lyl}. Hence, the LFVHD may
arise at one-loop level, as in the models of the standard seesaw,
inverse seesaw, and 3-3-1 model with massive neutrinos and inverse seesaw mechanism \cite%
{Pilaftsis:1992st,Korner:1992zk,Arganda:2004bz,Arganda:2014dta,Thao:2017qtn,Nguyen:2018rlb}. While the
standard seesaw model predicts suppressed branching ratios for LFVHD, these branchings can reach
interesting values of the order of $10^{-5}$ in the models with inverse seesaw mechanisms. Recent studies predict that
 the experimental sensitivities for LFVHD can reach values of the order of $10^{-5}$ in the near future~\cite{Chakraborty:2016gff,Qin:2017aju}.

The one-loop diagrams contributing to the LFV decays of $%
e_i\rightarrow e_j\ga$ and the SM-like Higgs boson decay $h^0_1\rightarrow e_i e_j$
with $i\neq j$ are exactly the same as those that appear in the seesaw and
inverse seesaw versions of the SM. The difference is the neutrino mixing matrix, arising from the linear seesaw mechanism. Hence, it will
be interesting to estimate how large the Br$(h^0_1\rightarrow e_ie_j)$
 can become under the current bounds of Br$(\mu\rightarrow e\ga)< 4.2 \times
10^{-13}$ \cite{TheMEG:2016wtm}.
It is expected that the future experimental sensitivities to the LFV decays will be improved, namely $6\times 10^{-14}$ for Br$(\mu\rightarrow\,e\ga)$~\cite{Baldini:2013ke,Baldini:2018nnn}, and about $\mathcal{O}(10^{-9})$ for the two decays Br$(\tau\rightarrow\,e\ga)$ and Br$(\tau\rightarrow\,\mu\ga)$~\cite{Kou:2018nap} (for a recent review see, for instance, Ref.~\cite{Calibbi:2017uvl}).

We will use the approximate formulas for the Br$(e_i\rightarrow e_j \ga)$ in 3-3-1 models
 given in Ref. \cite{Hue:2017lak}, which were checked to be well-consistent with the results obtained from the exact numerical computation. 
  Other approaches used for discussions of LFV decays of charged leptons in 3-3-1 models were also 
  given previously in the literature \cite{Boucenna:2015zwa, Arcadi:2017xbo,Lindner:2016bgg}. Analytic formulas for
calculating the one-loop contributions to LFVHD in the unitary
gauge are given in Ref. \cite{Thao:2017qtn, Nguyen:2018rlb, Hue:2015fbb},
 and were  shown to be consistent with previous works \cite%
{Arganda:2014dta}. Using these formulas, we only determine couplings between
physical states and ignore all Goldstone bosons.

From the  definition of the $SU(3)_L\times U(1)_X$ covariant derivative (\ref{eq:Covariant-Derivative}) we
 find its part related with the charged gauge bosons in our
model
\begin{align}  \label{Pcc}
\Pi^{CC}_\mu \equiv \fr 1 {\sqrt{2}} %
\begin{pmatrix}
0 & W^+_\mu  & 0 \\
W^-_\mu  & 0 & Y^-_\mu  \\
0 & Y^+_\mu  & 0%
\end{pmatrix}%
.
\end{align}
}
Hence the couplings of the SM-like Higgs with the charged gauge bosons are given by:
\begin{align}  \label{h01vv}
\mathcal{L}_{h^0_1VV}&\subset
\left(D_\mu \eta\right)^\+ \left(D^\mu \eta\right)
+\left(D_\mu \chi\right)^\+ \left(D^\mu \chi\right),  \crn
\mathcal{L}_{h^0_1VV}&=g\, m_W c_{\al}h^0_1 W^{+\mu} W^-_\mu  +  g\,m_Yh^0_1
s_{\al}Y^{+\mu} Y^-_\mu .
\end{align}
The matrix $U_{lL}$ in Eq. \eq{Ml} will be used to change the basis of
the left-handed charged leptons from the flavor basis to the physical one. Specifically, the correspondence between
 the original basis of the left-handed leptons and the physical one is  $\overline{e_L}R_{l L}\leftrightarrow 
  \overline{e_L}$, or $e_L\leftrightarrow R_{l L} e_L$, while the right handed ones are unchanged.
This means that $e_{iL} \to U_{lL,ij} e_{jL} $
and $\overline{e_{iL}} \to \overline{e_{jL}} U^\+  _{lL,ji}$ with $i,j=1,2,3$. 

  From Eqs.~\eqref{h01ee} and \eqref{h01vv}, we note that the couplings  SM-like Higgs boson with normal charged leptons and gauge boson  $W^\pm$   in the model under consideration ans the SM  are 
 $ (c_{\alpha} +s_{\alpha} t_{\theta})$ and $c_{\alpha}$, respectively. The lower bound $m_{Z'}\ge 4 $ TeV gives $v_{\chi}\ge 10$ TeV, which results in small $s_{\alpha}\simeq t_{2\alpha}/2 \sim  t_{\theta}\sim v/v_{\chi}\sim \mathcal{O}(10^{-2})$, therefore $c_{\alpha}=1+\mathcal{O}(10^{-4})$.  Similarly for the couplings of SM-like Higgs bosons with the SM quarks and the neutral gauge boson $Z$, where $\xi_{\eta}$   plays  role of the SM Higgs boson after the first breaking step. After the second one, the physical state of the SM-like Higgs boson is $ h^0_1 \simeq c_{\alpha} \xi_\eta$ and the relative difference  the Z boson with \Long{other} particle is $c_{\phi}$ with  $s_{\phi}\sim v^2/v^2_{\chi}$ given in Eq.~\eqref{eq_ZZpmix}. Hence, the largest relative differences between the couplings of the $h^0_1$ predicted by our model and the SM are $c_{\alpha}$ and $c_{\alpha}c_{\phi}$.  As a \Long{consequence,}
 these couplings of the SM-like Higgs bosons are still in the allowed regions constrained from experiments. 

The neutrino mass matrix $M_\nu$ in Eq. \eq{Mnu0} is diagonalized via
an unitary $9\times9$ matrix $U_\nu$, namely
\begin{align}  \label{define_Unu}
U_\nu^T M_\nu U_\nu&=\hat{M}_\nu=\mathrm{diag}(m_{n_1},m_{n_2},...,
\hat{m}_{n_9})=\mathrm{diag}(\hat{m}_\nu,\; \hat{m}_N ),
\end{align}
where $\hat{m}_\nu=\mathrm{diag}(m_{n_1},\;m_{n_2},\;m_{n_3})$ and $\hat{m}%
_N =\mathrm{diag}(m_{n_4},\;m_{n_5},...,\;m_{n_9})$ are the masses of active
and exotic neutrinos $n_L=(n_{1L},n_{2L},...,n_{9L})$. They are Majorana
fermions that satisfy $n_{kR}=n^c_{kL}$ with $k=1,2,...,9$. Relations
between the interaction and physical basis for the neutrino fields are: $(\overline{\nu^C_L}
\quad \overline{\nu_R} \quad \overline{N_R})=\overline{n_R}U^T_\nu$ and $%
(\nu_L \quad \nu^C_R \quad N^C_R)^T= U_\nu n_L$.

The couplings of charged gauge bosons with leptons are given by
\begin{align}  \label{lVenu}
\mathcal{L}_{V^{\pm}\ell\ell}&= i(\overline{L_L}\ga^\mu P^{CC}_{%
\mu}L_L)_1= \fr{g}{\sqrt{2}}\left( \overline{e_{iL}}\ga^\mu \nu_{iL}
W^-_\mu  +\overline{e_{iL}}\ga^\mu (\nu^c_{i})_L Y^-_\mu  +\mathrm{H.c.}%
\right).  \crn
\rightarrow \mathcal{L}_{V^{\pm}\ell\ell}&=\fr{g}{\sqrt{2}}\left[
(U_{lL})_{ji}(U_\nu)_{ ik} \overline{e_{jL}}\ga^\mu n_{kL} W^-_\mu
+(U_{lL})_{ji}(U_\nu)_{ (i+3)k} \overline{e_{jL}}\ga^\mu n_{kL}
Y^-_\mu  +\mathrm{H.c.}\right],
\end{align}
where the sums are taken for $i,j=1,2,3$ and $k=1,2,..,9$, and we have used $ (\nu^c_{i})_L=\nu^C_{iR}$.

Based on Eq. \eq{Lyl}, couplings of SM-like Higgs boson with neutrinos are
included in the following interactions:
\begin{align}  \label{h01nn}
-\mathcal{L}_{h^0_1nn} &\subset \left(1 +\fr{\xi_\eta }{v_\eta } +\fr{%
\xi_\chi }{v_\chi } \right)\left[ \overline{\nu^C_L}M_1 \nu^C_R +\mathrm{%
H.c.}\right] + \left(1 +\fr{\xi_\eta }{v_\eta } \right)\left[ \overline{%
\nu^C_L}M_2 N^C_R +\mathrm{H.c.}\right] +\left(1 +\fr{\xi_\chi }{v_\chi }
\right)\left[ \overline{\nu_R}M_3 N^C_R +\mathrm{H.c.}\right],  \crn
-\mathcal{L}_{h^0_1nn}& = \fr{g}{2 m_W}h^0_1 \left[ \left(c_{\al}
+s_{\al}t_\theta \right) \left( \overline{\nu^C_L}M_1 \nu^C_R +\mathrm{%
H.c.}\right) + c_{\al}\left( \overline{\nu^C_L}M_2 N^C_R +\mathrm{H.c.}%
\right) +s_\al t_\theta \left( \overline{\nu^C_L}M_3 N^C_R +\mathrm{%
H.c.}\right) \right]  \crn
&= \fr{gc_{\al}}{2 m_W}h^0_1 \left[ \left(1+ t_\al t_\theta
\right) (U_\nu)_{ik} (M_1)_{ij} (U_\nu)_{(j+3)p} + (U_\nu)_{ik}
(M_2)_{ij} (U_\nu)_{(j+6)p}\right.  \crn
&+\left. t_{\al}t_\theta (U_\nu)_{(i+3)k} (M_3)_{ij}
(U_\nu)_{(j+6)p}\right] \overline{n_{kR}} n_{pL} +\mathrm{H.c}.,
\end{align}
where the sums are taken for $i,j=1,2,3$ and $k,p=1,2,...,9$. By defining a
symmetric coefficient $\la_{kp}=\la_{pk}$  satisfying
\begin{equation*}
\la_{kp}\equiv \left(1+ t_{\al}t_\theta \right) (U_\nu)_{ik}
(M_1)_{ij} (U_\nu)_{(j+3)p} + (U_\nu)_{ik} (M_2)_{ij} (U_\nu)_{(j+6)p}
+ t_{\al}t_\theta (U_\nu)_{(i+3)k} (M_3)_{ij} (U_\nu)_{(j+6)p} + (k
\leftrightarrow p),
\end{equation*}
Eq. \eq{h01nn} can be written in the form
\be  \label{lh01nnl}
-\mathcal{L}_{h^0_1nn}= \fr{gc_\al}{4 m_W}h^0_1 \overline{n_k}\left[
\la_{kp} P_L + \la^*_{kp} P_R\right]n_p,
\ee
where $P_{L,R}=(1\mp \ga_5)/2$ are chiral operators and $n_{p,k}$ are
four-component spinors of Majorana neutrinos. This form of the couplings $h%
\overline{n_k}n_p$ allows us to use the Feynman rules in Ref.
\cite{Dreiner:2008tw} for calculating LFVHD at one loop level.

Based on Ref. \cite{Hue:2017lak}, the branching ratio for the $e_i\rightarrow e_j\ga$ ($i>j$) decay takes the form:

\begin{align}  \label{breijga}
\mathrm{Br}(e_i\rightarrow e_j\ga)&= \fr{12\pi^2}{G_F^2}|D_{ij}|^2\times \mathrm{Br}(e_i\rightarrow e_j\bar{\nu_j}\nu_i),
\end{align}
where $G_F=g^2/(4\sqrt{2}m_W^2)$ and $D_{ij}$ is the one-loop contribution
due to virtual charged gauge bosons and Majorana neutrinos running in the internal lines of the loops.
Such contribution can be written as $D_{ij}=D^{W}_{ij} +D^{Y}_{ij}$, where:
\begin{align}
D^W_{ij}&=-\fr{eg^2}{32\pi^2m_W^2} \sum_{k=1}^9%
\sum_{a,b=1}^3(U_{lL}^*)_{ib}
(U^*_\nu)_{bk}(U_{lL})_{ja}(U_\nu)_{ak}F(t_{kW}),  \crn
D^Y_{ij}&= -\fr{eg^2}{32\pi^2m_Y^2} \sum_{k=1}^9%
\sum_{a,b=1}^3 (U_{lL}^*)_{i(b+3)}
(U^*_\nu)_{(b+3)k}(U_{lL})_{j(a+3)}(U_\nu)_{(a+3)k}F(t_{kY}),
\end{align}
where
\begin{align}
t_{kW} \equiv \fr{m^2_{n_k}}{m_W^2},\; t_{kY}\equiv \fr{m^2_{n_k}}{m_Y^2}%
, \quad F(x) \equiv -\fr{10-43x+78x^2-49x^3 +4x^4 +18x^3\ln(x)}{12(x-1)^4}.
\end{align}
We note that $F(x)$ was given in Ref. \cite{Cheng:1980tp}. The above formulas
were used in the inverse seesaw 3-3-1 models \cite%
{Nguyen:2018rlb} and were confirmed to be numerically consistent with the
previous work of ~Re. \cite{Boucenna:2015zwa}. Numerical values of Br$(e_i\rightarrow e_j\bar{\nu_j}\nu_i)$ will
 be fixed as Br$(\mu\rightarrow e\bar{\nu_e}\nu_\mu)\simeq100\%$, Br$(\tau\rightarrow e\bar{\nu_e}
 \nu_\tau)\simeq17.82\%$, and Br$(\tau\rightarrow  \mu\bar{\nu_\mu}\nu_\tau)\simeq17.39\%$ \cite{Tanabashi:2018oca}.
 At low energy we take $g^2=e^2/s^2_W=4\pi\al_{\mathrm{em}}/s^2_W$, where $\al_{\mathrm{em}}\simeq1/137$
 and $s^2_W\simeq 0.231$.

\begin{figure}[t]
\centering
\includegraphics[width=0.8\textwidth]{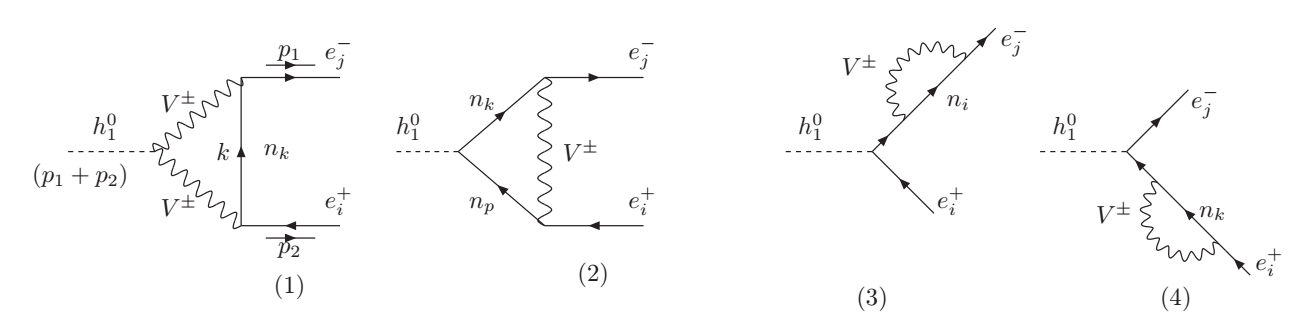}
\caption{ One-loop diagrams contributing to the SM-like Higgs boson decay $%
h^0_1\to e_ie_j$ in the unitary gauge, $V^{\pm}=W^{\pm},Y^{\pm}$.}
\label{figFeyn_LFVHD}
\end{figure}

For the LFVHD, one loop diagrams for Br$(h^0_1\to e_i e_j)$
are  shown in Fig.~\ref{figFeyn_LFVHD}.

The decay width for the process $h_1^0\rightarrow e_ie_j$ is given by:
\be
\Ga (h_1^0\rightarrow e_ie_j)\equiv\Ga (h_1^0\rightarrow e_i^{-}
e_j^{+})+\Ga (h^0_1\rightarrow e_i^{+} e_j^{-}) = \fr{ m_{h^0_1} }{8\pi
}\left(\vert \De_{(ij)L}\vert^2+\vert \De_{(ij)R}\vert^2\right),
\label{LFVwidth}
\ee
with the condition $m_{h_1^0}\gg m_{i,j}$ being $m_{i,j}$ the charged lepton
masses.

The corresponding branching ratio is
\bea\label{eq:Br-hee}
{\rm Br}(h^0_1\rightarrow e_ie_j)=
\Ga (h_1^0\rightarrow e_ie_j)/\Ga^{\mathrm{total}}_{h_1^0}
\eea
where $\Ga^{\mathrm{total}}_{h_1^0}\simeq 4.1\times 10^{-3}$ GeV \cite{Denner:2011mq}.
We define the $\De_{(ij)L,R}$ functions
\be
\De_{(ij)L,R} =\sum_{i=1}^4 \left(\De^{(i)W}_{(ij)L,R} +
\De^{(i)Y}_{(ij)L,R}\right),  \label{deLR}
\ee
where analytic forms for  the functions in the r.h.s. are shown in Appendix \ref{deltalr}
(for detailed calculations, see Refs. \cite{Thao:2017qtn,Hue:2015fbb}).
 The above formulas  were crosschecked using FORM~\cite{Vermaseren:2000nd,Kuipers:2012rf}.

Numerical input parameters we use for the analysis of the LFV processes correspond to the benchmark point given in Eq.~\eq{Benchmarklepton}, which implies that the corresponding values of the physical
observables of the lepton sector
 are automatically consistent with the neutrino oscillation experimental data.
 The mixing matrix of the charged lepton sector is fixed as given in Eq.~\eq{UlUnu}. The neutrino mixing matrix $U_\nu$ and neutrino masses
  can be numerically determined from Eq.~(\ref{define_Unu}), by using the numerical parameters given in~\eq{Benchmarklepton}.
According to our estimates $M^{(1)}_\nu$ is nearly independent of $v_\chi$. On the other hand the heavy neutrino masses show
 significant $v_\chi$-dependence, because they get main contributions from $M_3$ given in Eq. \eq{M123}.
 Furthermore they are nearly degenerate, which implies, $m_{n_4}\simeq m_{n_5}\simeq...\simeq\,m_{n_9}\simeq y^{(L)}_\chi\fr{v_\chi}{\sqrt{2}}\left( \fr{v_\si}\La \right) $ as indicated by Eqs.~\eq{M2nu} and \eq{M3nu}.  Hence we can see the dependence of the LFV branching ratios on the heavy neutrino masses, which are related to $v_{\chi}$ as shown by Eqs.~\eq{M2nu}, \eq{M3nu} and \eq{M123}. Besides the two VEVs $v_\phi$ and $v_\rho$ that were fixed in the discussion of the charged lepton sector, we choose $v_\xi=v_\si=v_\phi=\la \La$, while the three factors in front of the matrices $M_{1,2,3}$ in~Eq. \eq{M123} can be written in terms of $y_{1,2}$ as follows
	\begin{align}
	\label{eq_y12}
	y_1v_\eta \la^{16}&\equiv \fr{v_\eta v_\chi v_\zeta }{2\sqrt{2}\La^2}\left(
	\fr{v_\si }\La \right) ^{11}, \quad y_{1\eta }^{( L) }\fr{
		v_\eta v_\xi  }{\sqrt{6}\La }\left( \fr{ v_\si } \La \right) \equiv y_2v_\eta \la^2, \quad y_\chi^{( L) }\fr{v_\chi }{\sqrt{2}}\left( \fr{%
		v_\si }\La \right) = \fr { y_1y_2 v^2_\eta \la^{18}}{m_\nu},
	\end{align}
	where $y_{1,2} \sim \mathcal{O}(1)$. In our numerical analysis  
	we fix $\La\simeq 96$ TeV, and the CP-even neutral Higgs mixing parameters are set as follows $s_\al=0$, $c_\al=1$.   
In addition, we consider values for the $Z'$ mass satisfying $ M_{Z'} >4 $ TeV, which correspond to a $SU(3)_L\times U(1)_X$ symmetry breaking scale fulfilling
 $v_\chi>10$ TeV, as derived from the approximate formula $M^2_{Z'}\simeq g^2c^2_Wv^2_{\chi}/(3-4s^2_W) $~\cite{Long:2018dun}.  Numerical results for Br$(e_i\rightarrow e_j\ga)$ and  Br$(h^0_1\rightarrow e_ie_j)$ depending on $y_1$ and $y_2$  are illustrated in Table~\ref{table_LFVbr} for $v_\chi=15$ TeV. For $v_\chi$ around this value, all numerical results are the same hence it is unnecessary to discuss them here.
\begin{table}[ht]
	\centering
	\begin{tabular}{|c|c|c|c|c|c|c|c|}
		\hline
	$(y_1,y_2)$	&$m_{n_4}$ [GeV]& Br$(\mu\rightarrow e\ga)$ &  Br$(\tau\rightarrow e\ga)$& Br$(\tau\rightarrow \mu\ga)$  & Br$(h^0_1\rightarrow \mu\,e)$ & Br$(h^0_1\rightarrow \tau\,e)$& Br$(h^0_1\rightarrow \tau\mu)$\\
	\hline
	$(0.1,0.1)$	& $81.4$ & $2.8 \times 10^{-13}$  &  $8.5 \times 10^{-14}$ &  $6.8 \times 10^{-13}$& $3.1 \times10^{-18}$ & $4.4 \times10^{-13}$ &  $3.6 \times 10^{-12}$\\
	\hline
	$(0.5,0.1 )$	&  $407$  & $4.3 \times 10^{-15}$ &  $1.3  \times10^{-15}$ &  $1.1 \times 10^{-14}$ & $9.6  \times10^{-20}$ & $1.3  \times 10^{-14}$ &  $1.1  \times10^{-13}$ \\
	\hline
		$(2,0.1 )$	&  $1627.8$ & $2.5 \times 10^{-17}$ &   $7.8 \times 10^{-18}$ & $6.2 \times 10^{-17}$ & $1.2\times 10^{-19}$ & $1.6\times 10^{-14}$ &  $1.3\times 10^{-13}$ \\
	\hline
	$(5,0.1 )$	& $4069.5 $ &  $6.8\times 10^{-19}$&  $2.1\times 10^{-19}$ & $1.7\times 10^{-18}$ & $8.3\times 10^{-19}$ & $1.2\times 10^{-13}$ & $9.5\times 10^{-13}$ \\
	\hline
		$(0.1,0.5 )$	&  $4.1 \times 10^2$& $2.7\times 10^{-12}$ &   $8.2\times 10^{-13}$ & $6.6\times 10^{-12}$ & $6.1\times 10^{-17}$ &$8.5\times 10^{-12}$ &  $7.0\times 10^{-11}$\\
	\hline
	$(0.5, 0.5)$	&  $2.03 \times 10^3$& $6.6\times 10^{-15}$ & $ 2.0\times 10^{-15}$  & $ 1.6\times 10^{-14}$   & $ 1.2\times 10^{-16}$ & $ 1.6\times 10^{-11}$ & $1.4\times 10^{-10} $ \\
		\hline
	$(2, 0.5)$	&  $8.14 \times 10^3$& $2.7\times 10^{-17} $ &$ 8.2\times 10^{-18}$   &$ 6.6\times 10^{-17}$  & $ 1.6\times 10^{-15}$ & $ 2.3\times 10^{-10}$ & $1.9\times 10^{-9} $ \\
	\hline
	$(0.1, 2)$	& $ 1.63 \times 10^3$ & $ 4.1\times 10^{-12}$ & $1.2\times 10^{-12} $  & $9.9\times 10^{-12} $ & $ 1.9\times 10^{-14}$& $ 2.6\times 10^{-9}$& $2.2\times 10^{-8 }$ \\
	\hline
		$(0.5, 2)$	& $8.14 \times 10^3$ & $ 6.9\times 10^{-15}$ & $2.1\times 10^{-15} $  &$ 1.7\times 10^{-14}$  &$ 4.1\times 10^{-13}$ & $ 5.8\times 10^{-8}$& $4.8\times 10^{-7} $ \\
	\hline
			$(0.1, 4.5)$	&  $ 3.66 \times 10^3$& $ 4.2\times 10^{-12}$ & $ 1.3\times 10^{-12}$  &$ 1.0\times 10^{-11}$  &$2.8\times 10^{-12} $ &$3.9\times 10^{-7} $ &$ 3.2\times 10^{-6}$  \\
	\hline
	$(0.2,\; 4.5)$	& $7325.4$ & $2.7 \times 10^{-13}$ & $8.2 \times10^{-14}$  & $6.6 \times10^{-13}$ & $9.2 \times10^{-12}$ & $1.3 \times10^{-6}$& $1.1 \times 10^{-5}$  \\
	\hline
\end{tabular}
	\caption{Branching ratios for the LFV decays with $v_{\chi}=15$ TeV. The second column presents the numerical values of the heavy neutrino masses.}\label{table_LFVbr}
\end{table}
 The product $y_1y_2$ is constrained by the perturbative limit of the Yukawa coupling $ y_1y_2 \sim y^L_\chi<\sqrt{4\pi}\simeq 3.5$,  
 as follows from Eq.~\eq{eq_y12}. Table~\ref{table_LFVbr} shows the numerical values of the Branching ratios for the LFV decays
  with $v_{\chi}=15$ TeV and different values of the Yukawa couplings $y_1$ and $y_2$ and heavy neutrino masses.
 Notice that a specific value of $(y_1,y_2)$ in Table~\ref{table_LFVbr}, will predict a value for the Yukawa
  coupling $y^L_\chi \simeq \sqrt{2}m_{n_4}/(v_\chi \la) \leq 3.5$, leading to $m_{n_4}\leq 0.557 v_\chi$. Thus for $v_\chi=15$ TeV we have $m_{n_4}\leq 8.35$ TeV.

Based on the numerical results reported in Table~\ref{table_LFVbr}, we can see that Br$(\mu\rightarrow\,e\ga)$ can reach values close to its recent experimental bound provided that $y_1$ is small enough. On the other hand, Br$(h^0_1\rightarrow\mu\tau)$ can reach $\mathcal{O}(10^{-5})$ values when $y_2$ is large enough, like for example $y_2=4.5$ as shown in Table~\ref{table_LFVbr}. Furthermore, increasing $y_2$ will result in larger values for Br$(h^0_1\rightarrow\mu\tau)$. We can see  that the Br$(h^0_1\rightarrow e_ie_j)$ is enhanced  
when the heavy neutrino mass $m_{n_4}$ is increased, which is a generic behavior
observed in inverse seesaw models~\cite{Arganda:2014dta, Thao:2017qtn}. Because the experiment data favors  
lower bounds of $y_1$, and the perturbative limit of $y^L_\chi$  and $v_\chi$ results in upper bounds of $y_2$,  there exist upper bounds, which are order of $\mathcal{O}(10^{-5})$ and $\mathcal{O}(10^{-6})$ for the Branching ratios 
of the two decays $h^0_1\rightarrow \mu\tau,e\tau$ for the numerical values of the free parameters chosen above. The remaining LFV decays $\tau\rightarrow \mu\ga,e\ga$ and $h^0_1\rightarrow e\mu$ have much smaller Branching ratios than the characteristic sensitivities of current experimental searches.

\section{Conclusions}
\label{conclusion}
We constructed a viable multiscalar singlet extension of the 3-3-1 model with two scalar triplets and three right handed Majorana neutrinos  
where the tiny masses for the light active neutrinos are produced by the linear seesaw mechanism. Our model is based on the $A_4$ family symmetry, 
which is supplemented by other auxiliary symmetries. The observed pattern of the SM charged fermion masses and fermionic mixing parameters originates from the spontaneous breaking of the discrete symmetries of the model and does not require any fine-tuning of the model parameters.

We analyzed the implications of our model in the lepton flavour violating processes.
We demonstrated that the branching ratio Br$(\mu\rightarrow\,e\ga)$ can reach values close to the recent upper experimental bounds, thus constraining 
the values of Br$(\tau \rightarrow \mu\ga)$ and Br$(\tau \rightarrow e\ga)$ to
be much smaller than the corresponding experimental sensitivities.
On the other hand, the model allows
Br$(h^0_1\rightarrow \mu\tau)$ and Br$(h^0_1\rightarrow \,e\tau)$
to reach the values of about
$\mathcal{O}(10^{-5})$ and $\mathcal{O}(10^{-6})$, respectively. Besides that, we have studied the implications of our model in meson oscillations and we have found that our model is consistent with the constraints arising from meson mixings.
We also studied the production of the heavy $Z^{\prime }$ gauge boson
in proton-proton collisions via the Drell-Yan mechanism. We found that the corresponding total cross section ranges at the LHC from $0.11$ fb up 
to $0.01$ fb when the $Z^\prime $ gauge boson mass is varied within $7-8$ TeV interval. The $Z^\prime $ production cross section will be
significantly enhanced at the proposed energy upgrade of the LHC with $\sqrt{S}=28$ TeV reaching typical values of $82-30$ fb.
From these results we found that the $pp\to Z^\prime\to l^+ l^-$ resonant production cross section reach the values of about $10^{-3}$ fb and $1$ fb for $M_{Z^\prime }=7$ TeV at the energies
 $\protect\sqrt{s}=13$ TeV
and $\protect\sqrt{s}=28$ TeV, respectively.

The first value of the resonant production cross section is below and the second lies on the verge of the sensitivities of the LHC experiments at the corresponding energies.

\section*{Acknowledgments}

This research has received funding from ANID-Chile FONDECYT No. 1210378, No. 1190845, CONICYT PIA/Basal FB0821, Milenio-ANID-ICN2019\_044, the Vietnam National Foundation for Science and
Technology Development (NAFOSTED) under grant number 103.01-2019.387. A.E.C.H is very grateful to the Institute of Physics, Vietnam Academy of
Science and Technology for the warm hospitality and for financing his visit where this work was started.
\appendix

\section{The product rules for $A_4$}
\label{A4}
The $A_4$ group has one three-dimensional $\mathbf{3}$\ and
three distinct one-dimensional $\mathbf{1}$, $\mathbf{1}^\prime $ and $%
\mathbf{1}^{\prime \prime }$ irreducible representations, satisfying the
following product rules:
\bea
&&\hspace{18mm}\mathbf{3}\otimes \mathbf{3}=\mathbf{3}_{s}\oplus \mathbf{3}%
_a\oplus \mathbf{1}\oplus \mathbf{1}^\prime \oplus \mathbf{1}^{\prime
\prime },  \label{A4-singlet-multiplication} \\[0.12in]
&&\mathbf{1}\otimes \mathbf{1}=\mathbf{1},\hspace{5mm}\mathbf{1}^{\prime
}\otimes \mathbf{1}^{\prime \prime }=\mathbf{1},\hspace{5mm}\mathbf{1}%
^\prime \otimes \mathbf{1}^\prime =\mathbf{1}^{\prime \prime },\hspace{%
5mm}\mathbf{1}^{\prime \prime }\otimes \mathbf{1}^{\prime \prime }=\mathbf{1}%
^\prime ,  \notag
\eea%
Considering $\left( x_1 ,x_2,x_3 \right) $ and $\left(
y_1 ,y_2,y_3 \right) $ as the basis vectors for two $A_4$-triplets $%
\mathbf{3}$, the following relations are  fulfilled:

\bea
&&\left( \mathbf{3}\otimes \mathbf{3}\right) _{\mathbf{1}%
}=x_1 y_1 +x_2y_2+x_3 y_3 ,  \label{triplet-vectors} \\
&&\left( \mathbf{3}\otimes \mathbf{3}\right) _{\mathbf{3}_s}=\left(
x_2y_3 +x_3 y_2,x_3 y_1 +x_1 y_3 ,x_1 y_2+x_2y_1 \right) ,\
\ \ \ \left( \mathbf{3}\otimes \mathbf{3}\right) _{\mathbf{1}^\prime
}=x_1 y_1 +\om x_2y_2+\om^2x_3 y_3 ,  \crn
&&\left( \mathbf{3}\otimes \mathbf{3}\right) _{\mathbf{3}_a}=\left(
x_2y_3 -x_3 y_2,x_3 y_1 -x_1 y_3 ,x_1 y_2-x_2y_1 \right) ,\
\ \ \left( \mathbf{3}\otimes \mathbf{3}\right) _{\mathbf{1}^{\prime \prime
}}=x_1 y_1 +\om ^2x_2y_2+\om x_3 y_3 ,  \nn
\eea%
where $\om =e^{i\fr{2\pi }{3}}$. The representation $\mathbf{1}$ is
trivial, while the non-trivial $\mathbf{1}^\prime $ and $\mathbf{1}%
^{\prime \prime }$ are complex conjugate to each other. Some reviews of
discrete symmetries in particle physics are found in Refs. \cite%
{Ishimori:2010au,Altarelli:2010gt,King:2013eh, King:2014nza}. The discrete symmetry $A_4$ was first
implemented
to the 3-3-1 models in the Refs \cite{Yin:2007rv} and  \cite{Dong:2010gk}.

\section{Scalar sector}

\label{appHiggs}
Here we present more details about the scalar sector of our model containing the SM Higgs boson.

The scalar potential of the model can be splitted in the following two parts:
\begin{align}
		\label{Hpotential0}
		V_{S}=V^{\mathrm{invariant}}_{S} +V^{\mathrm{soft}}_{S}.
\end{align}
The first part $V^{\mathrm{invariant}}_{S}$ is invariant under the
$A_4\times Z_8\times Z_{14}\times Z_{22}$ discrete and $SU(3)_C\times SU(3)_L\times U(1)_X$ 
gauge symmetries,
	\begin{align}
V^{\mathrm{invariant}}_{S}&=\mu_\chi ^2 \chi^\dag\chi+\mu_\eta ^2 \eta^\dag\eta+
	\mu_{\si}^2 \si^*\si +\mu_\xi ^2(\xi^*\xi)_1  +
	\mu_\zeta ^2(\zeta^*\zeta)_1 +\mu_{\rho} ^2(\rho^*\rho)_1 +\mu_\va ^2(\va^*\va)_1  +\mu_\phi^2(\phi^*\phi)_1
 + \left[  \mu_{\phi\rho} ^2  (\phi^*\rho)_1 +\mathrm{H.c.}\right]
	\crn& + \la_\chi  (\chi^\dag\chi)^2 +
	\la_\eta  (\eta^\dag\eta)^2 + \la_{\si}(\si^*\si)^2  +  \sum_{S_i,S_j}\left[(S_i^*S_i)(S_j^*S_j)\right]_1
	\crn&+ \la_{\chi\eta} (\chi^\dag\chi)(\eta^\dag\eta) + \la^\prime
	_{\chi\eta} (\chi^\dag\eta) (\eta^\dag\chi)+ (\si^\ast\si) \left[ \la_{\chi\si}(\chi^\dag\chi)+ \la_{\eta\si}(\eta^\dag\eta) \right]
	\crn&+ \left\{ \left[(\rho\rho) (\phi^*\phi^*)\right]_1   +  \left[(\xi\rho) (\va ^*\va ^*)\right]_1  + \left[(\xi\phi) (\va ^*\va ^*)\right]_1   + \mathrm{H.c.} \right\}
	\crn&+ (\si^\ast\si) \left\{
	\sum_{S} \la_{S\si} (S^*S)_1
	+\left[ \la_{\phi\rho\si}   (\phi^*\rho)_1 +\mathrm{H.c.} \right]\right\}
	\crn&+ (\chi^\+  \chi) \left\{
	\sum_{S} \la_{S\chi}(S^*S)_1
	+\left[  \la_{\phi\rho\chi}   (\phi^*\rho)_1 +\mathrm{H.c.} \right]\right\}
	\crn&+ (\eta^\+  \eta)
	\sum_{S} \la_{S\eta}(S^*S)_1
	+\left[ \la_{\phi\rho\eta}   (\phi^*\rho)_1 +\mathrm{H.c.} \right]
	\crn&+
	\sum_{S}\left\{  \left[(\phi^*\rho)(S^*S)\right]_1 +\mathrm{H.c.}\right\},
	\label{Hpotential}
	\end{align}
	where $S,S_i,S_j=\xi,\zeta,\rho,\va,\phi$ are the scalar fields defined in Eq. (\ref{eq:scalars-def}).  The second part $V^{\mathrm{soft}}_{S}$ consists of $A_4\times Z_8\times Z_{14}\times Z_{22}$ soft-breaking terms needed to generate non-zero masses for the CP-odd neutral Higgs bosons as well as to solve the domain wall problem. 
%
%
The complete set of these soft-breaking terms is
	\begin{align}
	\label{eq_Vsoft}
	V^{\mathrm{soft}}_{S}&=	\mu'^2_{\sigma} \sigma^2 +  f_{\sigma} \sigma^3 + \sum_{S} \left[\mu^2_{1',S} \left( S^2\right)_{1'} + f_S\left( S^2\right)_{1}\sigma \right] +\mathrm{H.c.},
	\end{align}
where $S=\xi,\zeta,\rho,\va,\phi$; all  parameters  $\mu'_{\sigma}$, $f_{\sigma}$, $\mu^2_{1',S}$, and $f_S$ have the same dimension of mass.

The $A_4$-invariant products of four $A_4$-triplets $x,y,z,t$ can be decomposed as:

	\begin{align}
	\label{eq_A4xyzt}
	\left[ (xy)(zt)\right]_1&\equiv \la^{xyzt}_1 (xy)_1(zt)_1 + \la^{xyzt}_2 (xy)_{1'}(zt)_{1''} + \la^{xyzt}_3 (xy)_{1''}(zt)_{1'}  +\la^{xyzt}_4 \left[ (xy)_{3s}(zt)_{3s}\right]_1 \crn%
	&+\la^{xyzt}_5 \left[ (xy)_{3s}(zt)_{3a}\right]_1 +\la^{xyzt}_6 \left[ (xy)_{3a}(zt)_{3s}\right]_1 +\la^{xyzt}_7 \left[ (xy)_{3a}(zt)_{3a}\right]_1.
	\end{align}
	The products like $\left[ (xz)(yt)\right]_1$, $\left[ (xt)(yz)\right]_1$,... are not included in the scalar potential (\ref{Hpotential}) because they can always be written as linear combinations of the seven $A_4$-products in the right hand side of Eq.~(\ref{eq_A4xyzt}). This fact can be easily demonstrated, using the rules given in Appendix \ref{A4}.
	Let us note that due to the antisymmetry and symmetry properties of the $3_a$
	and $3_s$ triplet components in the products $(\xi\xi)$ and $(\zeta\zeta)$,
	we obtain $(3_s3_a)_1 +\mathrm{H.c.}=0$. Hence many terms of this kind
    does not appear in the scalar potential. Therefore, particular cases are written as
	\begin{align}\label{eq_A4SSSS}
	\left[(S^*S)^2\right]_1 &\equiv  \la^S _1(S^*S)_1(S^*S)_1 +
	\la^S _2(S^*S)_{1^\prime }(S^*S)_{1^{\prime \prime }}
	+\la^S _4 \left[(S^*S)_{3_s}(S^*S)_{3_s}\right]_1 +\la^S _7 \left[  (S^*S)_{3_a}(S^*S)_{3_a} \right]_1 ,
	\crn
	\left[S_i^*S_iS_j^*S_j\right]_1 &\equiv
	\la^{S_iS_j}_1(S_i^*S_i)_1 (S_j^*S_j)_1 +\left[\la^{S_iS_j}_2(S_i^*S_i)_{1^\prime }(S_j^*S_j)_{1^{\prime \prime }}+
	\mathrm{H.c.} \right] +\la^{S_iS_j}_4(S_i^*S_i)_{3_s}(S_j^*S_j)_{3_s}
	\crn &+\la^{S_iS_j}_7(S_i^*S_i)_{3_a}(S_j^*S_j)_{3_a}, \hs  S_i\ne S_j,
	\crn \left[(\xi\xi)(\va^*\va^*)\right]_1 &= \la'^{\xi\va}_1  (\xi\xi)_1(\va^*\va^*)_1  +\la'^{\xi\va}_2 (\xi\xi)_{1'}(\va^*\va^*)_{1''} +\la'^{\xi\va}_3 (\xi\xi)_{1''}(\va^*\va^*)_{1'}  +\la'^{\xi\va}_4 \left[(\xi\xi)_{3s}(\va^*\va^*)_{3s}\right]_1 ,
	\crn \left[(\rho\rho) (\phi^*\phi^*)\right]_1   &=\la'^{\rho\phi}_1  (\rho\rho)_1(\phi^*\phi^*)_1  +\la'^{\rho\phi}_2  (\rho\rho)_{1'}(\phi^*\phi^*)_{1''} +\la'^{\rho\phi}_3  (\rho\rho)_{1''}(\phi^*\phi^*)_{1'} +\la'^{\rho\phi}_4  \left[(\rho\rho)_{3s}(\phi^*\phi^*)_{3s} \right]_1 ,
	\crn  \left[(\xi\va^*)(\rho\phi^*)\right]_1    &= \la^{\xi\va\rho\phi^*}_1 (\xi\va^*)_1(\rho\phi^*)_1 + \la^{\xi\va\rho\phi^*}_2 (\xi\va^*)_{1'}(\rho\phi^*)_{1''} + \la^{\xi\va\rho\phi^*}_3 (\xi\va^*)_{1''}(\rho\phi^*)_{1'}  +\la^{\xi\va\rho\phi^*}_4 \left[ (\xi\va^*)_{3s}(\rho\phi^*)_{3s}\right]_1 \crn%
	&+\la^{\xi\va\rho\phi^*}_5 \left[ (\xi\va^*)_{3s}(\rho\phi^*)_{3a}\right]_1 +\la^{\xi\va\rho\phi^*}_6 \left[ (\xi\va^*)_{3a}(\rho\phi^*)_{3s}\right]_1 +\la^{\xi\va\rho\phi^*}_7 \left[ (\xi\va^*)_{3a}(\rho\phi^*)_{3a}\right]_1,
	\crn  \left[(\xi\va^*)(\rho^*\phi) \right]_1   &=\la^{\xi\va\rho^*\phi}_1 (\xi\va^*)_1(\rho^*\phi)_1 + \la^{\xi\va\rho^*\phi}_2 (\xi\va^*)_{1'}(\rho^*\phi)_{1''} + \la^{\xi\va\rho^*\phi}_3 (\xi\va^*)_{1''}(\rho^*\phi)_{1'}  +\la^{\xi\va\rho^*\phi}_4 \left[ (\xi\va^*)_{3s}(\rho^*\phi)_{3s}\right]_1 \crn%
	&+\la^{\xi\va\rho^*\phi}_5 \left[ (\xi\va^*)_{3s}(\rho^*\phi)_{3a}\right]_1 +\la^{\xi\va\rho^*\phi}_6 \left[ (\xi\va^*)_{3a}(\rho^*\phi)_{3s}\right]_1 +\la^{\xi\va\rho^*\phi}_7 \left[ (\xi\va^*)_{3a}(\rho^*\phi)_{3a}\right]_1.
	\end{align}
The above scalar potential has a fairly large number of scalar self-interactions.

The VEV's chosen in Eq.~\eq{VEVpattern} must  satisfy all the minimization
conditions of the scalar potential (\ref{Hpotential}), namely
\be\label{eq_mindHigg}
\left.\fr {\partial\,V_H}{\partial\,S^0}\right|_{S^0=\langle S^0\rangle,\forall\,S^0}=0.
\ee
The model contains 20 neutral scalar components, where three of them have zero VEVs. This  leads to  20 minimization equations relating the VEVs to the parameters of the scalar potential. We find that two equations for $\chi^0_1$ and $\rho^0_3$ are automatically satisfied.
The remaining 18 equations allow expressing 18 parameters of the model in terms of the other ones.

In order to generate fermions masses consistent with experiments we introduced in \eqref{VEVsinglets} the VEV pattern implying new relations between VEVs.
Let us show that this pattern is consistent with the scalar potential (\ref{Hpotential}).
It suffices to consider the simplified case of the scalar potential in the decoupling limit, when
the quartic couplings of the scalar $SU(3)_L$-triplets vanish, with the exception of two $ SU (3) _L $-triples. We will comment on more general cases later. The minimization conditions for the neutral scalars with real vev's   take  in the decoupling limit  the form

\begin{align}\label{eq_minEqHiggs}
S^0=\chi^0_3\rightarrow \mu _{\chi }^2& = -\fr {\la  _{\chi \eta } v_{\eta}^2}{2}-\la  _{\chi } v_{\chi }^2,\crn
S^0=\eta^0_1\rightarrow\mu _{\eta }^2&= -\fr {\la  _{\chi \eta } v_{\chi}^2}{2}-\la  _{\eta } v_{\eta }^2,\crn
S^0=\si ^0\rightarrow\mu _{\si  }^2&= - v_{\si  }^2 \la  _{\si  }  -2 \mu'^{2}_{\sigma }  -\frac{3 f_{\sigma } v_{\sigma }}{\sqrt{2}} -\frac{\sqrt{2} \left(f_{\zeta } v_{\zeta }^2+f_{\xi } v_{\xi }^2+f_{\rho} v_{\rho }^2\right)}{v_{\sigma }} 
\crn&\quad -\frac{\sqrt{2} f_{\phi } v_{\phi }^2 e^{i\psi}s_{2\alpha}}{ v_{\sigma }} +\frac{\sqrt{2} f_{\varphi } v_{\varphi }^2 e^{-i\psi}s_{2\alpha}}{ v_{\sigma }}
,\crn
S^0=\xi_1^0,\xi_2^0,\xi_3^0\rightarrow\mu _{\xi }^2&= -\sqrt{2} f_{\xi } v_{\sigma } -\fr {2}{3} v_{\xi}^2 (3 \la  _1^{\xi }+4 \la  _4^{\xi }),
\crn \mu _{1'\xi }^2&= 0, 
\crn S^0=\zeta_1^0,\zeta_3^0\rightarrow\mu _{\zeta }^2&= -\sqrt{2} f_{\zeta } v_{\sigma } -v_\zeta ^2 (2 \la  _1^{\zeta }+\la  _3^{\zeta }+2 \la  _4^{\zeta }),\crn
%
%
\mu _{1'\zeta }^2&= -\frac{v_{\zeta }^2 (2 w+1) (\lambda _2^{\zeta }-\lambda _3^{\zeta })}{2 (w+2)}, 
\crn S^0=\rho_1^0,\rho_2^0, \rho_3^0\rightarrow\mu _{\rho }^2&= -\fr {2}{3} v_\rho^2 (3 \la  _1^{\rho }+4 \la  _4^{\rho }), \crn
\mu _{\phi \rho }^{2}&= 0,\; \mu _{1'\rho }^2= 0,
\end{align}
where we have used that $\sum_{i=1}^3\langle \phi_i\rangle ^2=v_{\phi }^2 e^{i\psi}s_{2\alpha}$ and  $\sum_{i=1}^3\langle \varphi_i\rangle ^2= -v_{\varphi }^2 e^{-i\psi}s_{2\alpha}$.

Next, we consider the $A_4$-triplets $\phi$ and $\va $ with complex VEVs given in Eq.~\eqref{VEVpattern}. With $\mu _{\phi \rho }=0$,  we have 
three different minimization equations for $\phi$ in the following forms:

  \begin{align}
 	\label{eq_minPhi0}
 	S^0=\phi^0_1\rightarrow 0=& \fr {3 x_1^2 \mu _\phi^2}{2 v_\phi^2}+\la  _1^\phi
 	\left(x_1^2+x_2^2+x_3^2\right)+\la  _2^\phi \left(2 x_1^2-x_2^2-x_3^2\right)+4 x_1^2 \la  _4^\phi +\frac{3 \mu _{1'\phi }^2}{v_{\phi }^2} +\frac{3 f_{\phi } v_{\sigma }}{\sqrt{2} v_{\phi }^2},\crn
 	S^0=\phi^0_2\rightarrow 0=& \fr {3 x_2^2 \mu _\phi^2}{2 v_\phi^2}+\la  _1^\phi
 	\left(x_1^2+x_2^2+x_3^2\right)+\la  _2^\phi \left(-x_1^2+2 x_2^2-x_3^2\right)+4 x_2^2 \la  _4^\phi +\frac{3w \mu _{1'\phi }^2}{v_{\phi }^2} +\frac{3 f_{\phi } v_{\sigma }}{\sqrt{2} v_{\phi }^2},\crn
 	S^0=\phi^0_3\rightarrow 0=& \fr {3 x_3^2 \mu _\phi^2}{2 v_\phi^2}+\la  _1^\phi
 	\left(x_1^2+x_2^2+x_3^2\right)+\la  _2^\phi \left(-x_1^2-x_2^2+2 x_32\right)+4 x_3^2 \la  _4^\phi  +\frac{3w^2 \mu _{1'\phi }^2}{v_{\phi }^2} +\frac{3 f_{\phi } v_{\sigma }}{\sqrt{2} v_{\phi }^2},
 \end{align}
where $x_1=c_\al +e^{i\psi}s_{\al}$, $x_2=w \left(c_\al +we^{i\psi}s_\al\right)$, and $x_3=w^2 \left(c_\al +w^2e^{i\psi}s_\al\right)$ that satisfy $x_1+x_2+x_3=0$. Other relations used in our calculation are: $ \sum_{i=1}^3 x_i^2= 6c_{\alpha} s_{\alpha} e^{i\psi}$.    
From the scalar potential minimization equations we find: 

\begin{align} \label{eq_minPhi1}
\mu _\phi^2 &= -\fr 2 3 v_\phi^2 (3 \la  _1^\phi+4 \la  _4^\phi)  -\frac{3 \sqrt{2} f_{\phi } v_{\sigma }}{x_1^2+x_2^2+x_3^2}, \ %
\mu _{1'\phi }^2=0, \crn
\la_2^\phi&=\la_1^\phi + \frac{3 f_{\phi } v_{\sigma }}{\sqrt{2} v_{\phi }^2 \left(x_1^2+x_2^2+x_3^2\right)}. 
\end{align}

In the same way, we treat the minimization conditions for $\va $ and find the following relations
\begin{align} \label{eq_minvPhi}
\mu _\va  ^2 &= -\fr 2 3  v_\va^2 (3 \la  _1^\va  +4 \la  _4^\va  ) -\frac{3 \sqrt{2} f_{\varphi } v_{\sigma }}{y_1^2+y_2^2+y_3^2}, \; \mu _{1'\varphi }^2= 0,\crn
\la  _2^\va&= \la  _1^\va + \frac{3 f_{\varphi } v_{\sigma }}{\sqrt{2} v_{\varphi }^2 \left(y_1^2+y_2^2+y_3^2\right)},
\end{align}
where $y_1=c_{\alpha}-e^{-i\psi} s_{\alpha}$, $y_2=w^2\left( c_{\alpha}-w^2e^{-i\psi} s_{\alpha}\right)$, and $y_3=w\left( c_{\alpha}-we^{-i\psi} s_{\alpha}\right)$.

Thus, we see that the minimization conditions in the decoupling limit do not constrain the vev's.
This conclusion is valid in the general case, when all the quartic coupling return back to the scalar potential. This is trivially because these couplings just introduce new independent parameters, which can not introduce any constraint on the vev's.

Let us identify the SM-like Higgs boson with one of the scalars of our model or their linear combination.%

Note that the neutral CP-even components of the Higgs bosons always contain only one
massless state absorbed by the gauge boson $X^0$. This state is one of the
linear combinations of the two real components $R(\chi^0_1)$ and $%
R(\eta^0_3) $, which have zero VEVs.
More precisely, the model contains two would-be Goldstone bosons  $G_X, G^*_X$,  a neutral CP-odd Higgs boson $h_a$, and a mass eigenstate $h^0_3$. Namely, defining

\begin{equation*}
t_\theta=\tan\theta=\fr{v_\eta }{v_\chi },
\end{equation*}
we have the following relations between the original and the mass eigenstates of the neutral Higgs bosons
\begin{align}
\begin{pmatrix}
R_{\chi_1} \\
R_{\eta_3}%
\end{pmatrix}%
&=%
\begin{pmatrix}
c_\theta & s_\theta \\
-s_\theta & c_\theta%
\end{pmatrix}
\begin{pmatrix}
G_X \\
h^0_3%
\end{pmatrix},\; \begin{pmatrix}
I_{\chi_1} \\
I_{\eta_3}%
\end{pmatrix}%
=%
\begin{pmatrix}
c_\theta & s_\theta \\
-s_\theta & c_\theta%
\end{pmatrix}
\begin{pmatrix}
\overline{G}_X \\
h_a%
\end{pmatrix}
,  \crn
m_{G_X}&=m_{\overline{G}_X}=0,\quad m^2_{h^0_3}= m^2_{h_a}=\fr 1 2 \la^{\prime
}_{\eta\chi}(v^2_\eta  +v^2_\chi ).
\end{align}
The $R_{\zeta_2}$ is one mass eigenstate with mass $m^2_{R_{\zeta_2}}=\fr{2%
}{9}( -3\la) v_\zeta ^2$.

The remaining CP-even components of the neutral Higgs boson consist of  17
states $\xi_\chi  = \sqrt{2}R_{\chi^0_3}$, $\xi_\eta =\sqrt{2}R_{\eta^0_1}$, $R_{\si}$, $R_{\xi_i}$ ($i=1,2,3$), $R_{\zeta_1}$, and $R_{\zeta_3}$.
The squared mass matrix of these states is the $17\times17$ matrix denoted as $
\mathcal{M}^2_h$. This matrix has non-zero determinant, which means that all the
neutral CP-even Higgs bosons are massive. In  addition, $%
\mathrm{Det}[\mathcal{M}^2_h]_{v_\eta =0}=0$ implies that there is at least
one Higgs boson with mass at the electroweak scale. That lightest CP even scalar state is identified with the SM-like $126$ GeV Higgs boson.

To illustrate that there is one Higgs that can be identified with the
$126$ GeV SM-like Higgs boson found by LHC, we consider the simplified case when the
$SU(3)_L$ triplets $\chi$ and $\eta$ decouple from  $S=\si$, $\xi_i$, $\zeta_1,\; \zeta_3$, $\rho_i,\; \varphi_i$, and $\phi_i$ so that the corresponding quartic couplings vanish $\la_{\eta S}=\la_{\chi S}=0$. 
Then, the matrix $\mathcal{M}^2_h$ is split into two block-diagonal $2\times 2$ and $15\times 15$ matrices.
The first matrix in the basis $( \xi_\eta ,\;\xi_\chi )$ takes form

\begin{align}  \label{Hetachi}
\mathcal{M}^2_{h1}=%
\begin{pmatrix}
2\la_\eta  v_\eta ^2 & \la_{\eta\chi} v_\chi v_\eta  \\
\la_{\eta\chi} v_\chi v_\eta  & 2 \la_\chi  v_\chi ^2%
\end{pmatrix}%
.
\end{align}
Its mass eigenstates, $h^0_1$ and $h^0_2$, and their masses are
\begin{align}  \label{h012}
m^2_{h^0_{1,2}}&= \la_\chi  v_\chi ^2 + \la_\eta  v_\eta ^2 \mp
\sqrt{\left(\la_\chi  v_\chi ^2 -\la_\eta v_\eta ^2 \right)^2
+\la_{\eta\chi}^2 v_\chi ^2 v_\eta ^2},  \crn
\begin{pmatrix}
\xi_\eta  \\
\xi_\chi %
\end{pmatrix}%
&=
\begin{pmatrix}
c_{\al} & s_{\al} \\
-s_{\al} & c_{\al}%
\end{pmatrix}
\begin{pmatrix}
h^0_1 \\
h^0_2%
\end{pmatrix}%
,\quad t_{2\al}\equiv \tan(2\al)=\fr{\la_{\eta\chi}t_\theta}{%
\la_\chi  -\la_\eta  t_\theta^2}.
\end{align}
These two neutral Higgs bosons are similar in many respects to those discussed in the model \cite{Barreto:2017xix}.
Analogously to this model, in our case
in the limit $t_\theta \ll1$, we find that $m^2_{h^0_1}\simeq \left(
2\la_\eta  -\fr{\la_{\eta\chi}^2}{\la_\chi }\right)
v^2_\eta $, as should be for the SM Higgs boson, the mass of which is generated on the electroweak scale. Thus, we identify $h^{0}_{1}$  with the
SM-like Higgs boson found by the LHC. The simplified case when $t_\theta \ll1$ is used in our
discussion of the LFV Higgs decays in Sec. \ref{lfvdecay}.

The soft breaking terms introduced in the Higgs potential \eqref{eq_Vsoft} are enough to generate non-zero masses for all CP-odd Higgs bosons in the model under consideration, even some of them vanish by the minimization conditions of the Higgs potential. Namely, in the limit of $f_S=0$ with all $S=\sigma,\xi,\; \zeta, \rho, \; \varphi,\; \phi$, the total squared mass matrix of the CP-odd neutral components in the basis $S_0=(I_{\sigma},\;I_{\xi_i},\; I_{\zeta_i}, I_{\rho,_i}  \; I_{\phi_i},\; I_{\varphi_i} )$ separates into the six block sub-matrices, including one physical state $I_{\sigma}$ and another five $3\times 3$ matrices. 
\begin{align}
	\label{eq_CPoddmas}
	m^2_{a_\sigma}&=-4\mu'^{2} _{\sigma },\; 
\crn m^2_{\xi_i}&=\left\{\frac{1}{3} v_{\xi }^2 (3 \lambda _1^{\xi }+4 \lambda _4^{\xi }),\frac{1}{3} v_{\xi }^2 (-15 \lambda _1^{\xi }+18 \lambda _2^{\xi }+4 \lambda _4^{\xi }),\frac{1}{3} v_{\xi }^2 (-15 \lambda _1^{\xi }+18
\lambda _2^{\xi }+4 \lambda _4^{\xi })\right\},
\crn M^2_{I_\zeta}&=\left(
\begin{array}{ccc}
	\frac{-3 \lambda _3^{\zeta } w-4 \lambda _1^{\zeta } (w+2)+\lambda _2^{\zeta } (7 w+8)}{2 (w+2)} & 0 & 2 \lambda _1^{\zeta }-\lambda _2^{\zeta }-\lambda _3^{\zeta } \\
	0 & \frac{3 \lambda _2^{\zeta }+\lambda _3^{\zeta }+8 \lambda _4^{\zeta }+2 \lambda _3^{\zeta } w+4 \lambda _4^{\zeta } w-8 \lambda _1^{\zeta } (w+2)}{2 (w+2)} & 0 \\
	2 \lambda _1^{\zeta }-\lambda _2^{\zeta }-\lambda _3^{\zeta } & 0 & \frac{3 \lambda _3^{\zeta } (w+1)-4 \lambda _1^{\zeta } (w+2)+\lambda _2^{\zeta } (w+5)}{2 (w+2)} \\
\end{array}
\right)\AECH{v_{\zeta }^2},
\crn m^2_{a_{\rho_i}}&= \frac{v_{\rho^2}}{3}\times \{3 \lambda _1^{\rho }+4 \lambda _4^{\rho },-15 \lambda _1^{\rho }+18 \lambda _2^{\rho }+4 \lambda _4^{\rho },-15 \lambda _1^{\rho }+18 \lambda _2^{\rho }+4 \lambda _4^{\rho }\}, 
\crn 
 \frac{M^2_{I_{\phi}}}{ v^2_{\phi}}&=\mathrm{diag}\left\{2 \lambda _4^{\phi } \left( x_1^2+ x_2^2-\frac{4}{3}\right)+\left(3 x_3^2-2\right) \lambda _1^{\phi },\; 2 \lambda _4^{\phi } \left( x_1^2+ x_3^2 -\frac{4}{3}\right)+\left(3 x_2^2-2\right) \lambda
 _1^{\phi },\left(3 x_1^2-2\right) \lambda _1^{\phi }+ 2\lambda _4^{\phi } \left( x_2^2+ x_3^2-\frac{4}{3}\right)\right\},
 \crn \frac{M^2_{I_{\varphi}}}{v^2_{\varphi}} &=\mathrm{diag}\left\{2 \lambda _4^{\varphi } \left( y_1^2+ y_2^2 -\frac{4}{3}\right)+\left(3 y_3^2-2\right) \lambda _1^{\varphi },2 \lambda _4^{\varphi } \left(3 y_1^2+3 y_3^2-\frac{4}{3}\right)+\left(3
 y_2^2-2\right) \lambda _1^{\varphi },\left(3 y_1^2-2\right) \lambda _1^{\varphi } +2 \lambda _4^{\varphi } \left( y_2^2+ y_3^2 -\frac{4}{3}\right)\right\},
\end{align}
where $m^2_{a_{S_i}}$ denotes the squared mass eigenstate of the CP-odd Higgs boson corresponding to the original basis $\{I_{S_i}\}$.  It can be seen that the CP-odd Higgs boson masses get contributions from the discrete symmetry preserving terms. Other trilinear soft-breaking terms with $f_S\neq 0$ will yield complicated mixings among these Higgs bosons, without affecting the 
  phenomenology of our model, since this scalar sector, being very heavy, is decoupled from the SM fields.  Notice that since we are considering a CP conserving scalar potential,  the heavy neutral CP odd scalars do not mix with the CP even electrically neutral component of the $SU(3)_L$ scalar triplet $\eta$. On the other hand, the heavy physical scalar states arising from the gauge singlet scalars are mainly decoupled from the $126$ GeV SM like Higgs boson due to the very small mixings between the scalar singlets and the CP even electrically neutral component of $\eta$. Consequently, we are in the decoupling scenario where the coupling strengths of the $126$ GeV SM like Higgs boson with SM particle are very close to the SM expectation. In view of the above, 
 setting $f_S\neq 0$ will not affect the main physics results of this paper. One can also think about introduction of an additional ad hoc symmetry forbidding the trilinear terms in (\ref{eq_Vsoft}) and thus guarantying $f_{S}=0$. 
The study of this possibility goes beyond the scope of this paper and is deferred for a future work.

\section{\label{deltalr}Analytic formulas of LFVHD at the one loop level}

\allowdisplaybreaks
One-loop contributions to LFVHD defined in Eq. (\ref{deLR}) are written in
terms of Passarino-Veltman (PV) functions  \cite{Passarino:1978jh}. In
this work, they are denoted as $B^{(i)}_{0,1},\,B^{(12)}_0,\, C_0$ and $%
C_{1,2}$. In the limit $m_{i,j}\simeq0$, their analytic formulas were given
in Refs. \cite{Hue:2015fbb,Thao:2017qtn,Denner:2005nn}. These functions are
used for our numerical analysis.
It has been shown numerically that they are in a good agreement with
the exact results computed by LoopTools \cite{Hahn:1998yk} in Ref.~\cite{Phan:2016ouz}.

The analytic expressions of $\De^{(i)W}_{L,R}\equiv
\De^{(i)W}_{(ij)L,R}$ given in Eq.~\eqref{deLR}, where $i$ implies the diagram (i) in Fig. \ref%
{figFeyn_LFVHD}, are
\begin{align}
\De^{(1)W}_L &= -\fr{g^3c_{\al} m_{j}}{64\pi^2 m_W^3}%
\sum_{k=1}^{9}\sum_{a,b=1}^3 (U_\nu)_{ak}(U^*_\nu)_{bk} (U_{\ell
L})_{ja}(U^*_{\ell L})_{ib}  \crn
&\times \left\{ m_{n_k}^2\left(B^{(1)}_1- B^{(1)}_0- B^{(2)}_0\right) -m_i^2
B^{(2)}_1 +\left(2m_W^2+m^2_{h^0_1}\right)m_{n_k}^2 C_0 \right.  \crn
&-\left. \left[2m_W^2\left(2m_W^2+m_{n_k}^2+m_j^2-m_i^2\right) +
m_{n_k}^2m_{h^0_1}^2\right] C_1 + \left[2m_W^2\left(m_j^2-m^2_{h^0_1}%
\right)+ m_i^2 m^2_{h^0_1}\right]C_2\fr{}{}\right\},  \crn
\De^{(1)W}_R &= -\fr{g^3c_{\al} m_i}{64\pi^2 m_W^3}%
\sum_{k=1}^{9}\sum_{a,b=1}^3 (U_\nu)_{ak}(U^*_\nu)_{bk} (U_{\ell
L})_{ja}(U^*_{\ell L})_{ib}  \crn
&\times \left\{ -m_{n_k}^2\left(B^{(2)}_1+B^{(1)}_0+ B^{(2)}_0\right) +m_j^2
B^{(1)}_1 +\left(2m_W^2+m^2_{h^0_1}\right)m_{n_k}^2 C_0 \right.  \crn
&-\left. \left[2m_W^2\left(m_i^2-m^2_{h}\right)+ m_j^2 m^2_{h^0_1}\right]C_1
+ \left[2m_W^2\left(2m_W^2+m_{n_k}^2-m_j^2+m_i^2\right) +
m_{n_k}^2m_{h^0_1}^2\right] C_2 \fr{}{}\right\},  \notag  \label{de1wL} \\
\De^{(2)W}_L &= -\fr{g^3c_{\al} m_j}{64\pi^2 m_W^3}%
\sum_{k,p=1}^{9} \sum_{a,b=1}^3  (U_\nu)_{ak}(U^*_\nu)_{bp} (U_{\ell
L})_{ja}(U^*_{\ell L})_{ib}  \crn
&\times \left\{\la^{0*}_{kp}m_{n_p}\left[B^{(12)}_0-m_W^2C_0+\left(2
m_W^2+m_{n_k}^2-m_j^2\right)C_1\right]\right.  \crn
&\hspace{0.4cm}\left.+\la^{0}_{kp}m_{n_k}\left[B^{(1)}_1+\left(2
m_W^2+m_{n_p}^2-m_i^2\right)C_1\right]\right\},  \crn
\De^{(2)W}_R  &=-\fr{g^3c_{\al} m_i}{64\pi^2 m_W^3}\sum_{k,p=1}^{9}
\sum_{a,b=1}^3  (U_\nu)_{ak}(U^*_\nu)_{bp} (U_{\ell L})_{ja}(U^*_{\ell
L})_{ib}  \crn
&\times \left\{\la
_{kp}m_{n_k}\left[B^{(12)}_0-m_W^2C_0-\left(2
m_W^2+m_{n_p}^2-m_i^2\right)C_2\right]\right.  \crn
&\hspace{0.4cm}-\left.\la^{0*}_{kp}m_{n_p}\left[B^{(2)}_1+\left(2
m_W^2+m_{n_k}^2-m_j^2\right)C_2\right]\right\},  \crn
\De^{(3+4)W}_L &=- \fr{g^3m_jm_i^2(c_{\al} +s_\al t_\theta)}{64%
\pi^2m^3_W(m_j^2-m_i^2)} \sum_{k=1}^{9}\sum_{a,b=1}^3 (U_\nu)_{ak}(U^*_{%
\nu})_{bk} (U_{\ell L})_{ja}(U^*_{\ell L})_{ib}  \crn
&\times \left[ 2m_{n_k}^2\left(B^{(1)}_0-B^{(2)}_0\right) - \left(2 m_W^2
+m_{n_k}^2\right) \left(B^{(1)}_1 +B^{(2)}_1 \right)- m_j^2 B^{(1)}_1 -m_i^2
B^{(1)}_2 \right],  \crn
\De^{(3+4)W}_R  &=\fr{m_j}{m_i}\De^{(3+4)W}_L  \crn
\De^{(1)Y}_L &= -\fr{g^3s_{\al} m_{j}}{64\pi^2 m_Y^3}%
\sum_{k=1}^{9}\sum_{a,b=1}^3 (U_\nu)_{(a+3)k}(U^*_\nu)_{(b+3)k}
(U_{\ell L})_{ja}(U^*_{\ell L})_{ib}  \crn
&\times \left\{ m_{n_k}^2\left(B^{(1)}_1- B^{(1)}_0- B^{(2)}_0\right) -m_i^2
B^{(2)}_1 +\left(2m_Y^2+m^2_{h^0_1}\right)m_{n_k}^2 C_0 \right.  \crn
&-\left. \left[2m_Y^2\left(2m_Y^2+m_{n_k}^2+m_j^2-m_i^2\right) +
m_{n_k}^2m_{h^0_1}^2\right] C_1 + \left[2m_Y^2\left(m_j^2-m^2_{h^0_1}%
\right)+ m_i^2 m^2_{h^0_1}\right] C_2\fr{}{}\right\},  \crn
\De^{(1)Y}_R &= -\fr{g^3c_{\al} m_i}{64\pi^2 m_Y^3}%
\sum_{k=1}^{9}\sum_{a,b=1}^3 (U_\nu)_{(a+3)k}(U^*_\nu)_{(b+3)k}
(U_{\ell L})_{ja}(U^*_{\ell L})_{ib}  \crn
&\times \left\{ -m_{n_k}^2\left(B^{(2)}_1+B^{(1)}_0+ B^{(2)}_0\right) +m_j^2
B^{(1)}_1 +\left(2m_Y^2+m^2_{h^0_1}\right)m_{n_k}^2 C_0 \right.  \crn
&-\left. \left[2m_Y^2\left(m_i^2-m^2_{h}\right)+ m_j^2 m^2_{h^0_1}\right]C_1
+ \left[2m_Y^2\left(2m_Y^2+m_{n_k}^2-m_j^2+m_i^2\right) +
m_{n_k}^2m_{h^0_1}^2\right] C_2 \fr{}{}\right\},  \crn
\De^{(2)Y}_L &= -\fr{g^3c_{\al} m_j}{64\pi^2 m_Y^3}%
\sum_{k,p=1}^{9} \sum_{a,b=1}^3  (U_\nu)_{(a+3)k}(U^*_\nu)_{(b+3)p}
(U_{\ell L})_{ja}(U^*_{\ell L})_{ib}  \crn
&\times \left\{\la^{0*}_{kp}m_{n_p}\left[B^{(12)}_0-m_Y^2C_0+\left(2
m_Y^2+m_{n_k}^2-m_j^2\right)C_1\right]\right.  \crn
&\hspace{0.4cm}\left.+\la^{0}_{kp}m_{n_k}\left[B^{(1)}_1+\left(2
m_Y^2+m_{n_p}^2-m_b^2\right)C_1\right]\right\},  \crn
\De^{(2)Y}_R  &=-\fr{g^3c_\al m_i}{64\pi^2 m_Y^3}\sum_{k,p=1}^{9}
\sum_{a,b=1}^3  (U_\nu)_{(a+3)k}(U^*_\nu)_{(b+3)p} (U_{\ell
L})_{ja}(U^*_{\ell L})_{ib}  \crn
&\times \left\{\la^{0}_{kp}m_{n_k}\left[B^{(12)}_0-m_Y^2C_0-\left(2
m_Y^2+m_{n_p}^2-m_i^2\right)C_2\right]\right.  \crn
&\hspace{0.4cm}-\left.\la^{0*}_{kp}m_{n_p}\left[B^{(2)}_1+\left(2
m_Y^2+m_{n_k}^2-m_j^2\right)C_2\right]\right\},  \crn
\De^{(3+4)Y}_L &=- \fr{g^3m_jm_i^2(c_\al +s_\al  t_\theta)}{64%
\pi^2m^3_Y(m_j^2-m_i^2)} \sum_{k=1}^{9}\sum_{a,b=1}^3 (U_{%
\nu})_{(a+3)k}(U^*_\nu)_{(b+3)k} (U_{\ell L})_{ja}(U^*_{\ell L})_{ib}
\crn
&\times \left[ 2m_{n_k}^2\left(B^{(1)}_0-B^{(2)}_0\right) - \left(2 m_Y^2
+m_{n_k}^2\right) \left(B^{(1)}_1 +B^{(2)}_1 \right)- m_j^2 B^{(1)}_1 -m_i^2
B^{(1)}_2 \right],  \crn
\De^{(3+4)Y}_R  &=\fr{m_j}{m_i}\De^{(3+4)Y}_L.
\end{align}

\section{Couplings of the $Z$ and $Z^{\prime}$ gauge bosons to fermions}
\label{app_Zcoupling}
The interactions between fermions and neutral gauge bosons are determined as
\be
\mathcal{L}_{\mathrm{ngaugefermion}} = g \overline{f} \ga^\mu P_\mu^{NC} f \, ,
\label{eq197}
\ee
where $f$ denotes all fermions in the model under consideration. Then one gets

\bit
\item Electromagnetic interaction, as usual: $\mathcal{L}_{em} = e \overline{f}
\ga^\mu Q f A_\mu$ .

\item Interaction between $Z$ with fermion
\begin{align}
\mathcal{L}_{Z f}& = \fr{g}{c_W}\overline{f} \ga^\mu \left[ c_\phi \left( T_3
-s^2_W Q\right) - s_\phi \left( \fr{\sqrt{3-4s^2_W}}{\sqrt{3}} T_8 +\fr{
	s^2_W }{\sqrt{3-4s^2_W}} X\right)\right] f Z_\mu \crn
&\equiv   \fr{g}{c_W}\overline{f}_{L,R} \ga^\mu g_{L,R} f_{L,R}
Z_\mu \, , \label{eq198}
\end{align}
where $\phi$ is the $Z-Z'$ mixing angle  given in Ref~\cite{Long:2018dun}, $s_\phi \equiv \sin\phi$, $c_\phi \equiv \cos\phi$,
\begin{align}
\label{eq_ZZpmix}
\tan\phi\simeq 	s_\phi \simeq  \fr {(1-2s_W^2)\sqrt{3-4s_W^2}}{4 c_W^4} \left(\fr {v_{\eta}^2}{v^2_{\chi}}\right),\; M^2_{Z'}\simeq \fr {g^2c_W^2}{4(3-4s_W^2)} \left[ 4v^2_{\chi} +\fr {v^2_{\eta}(1-2s^2_W)^2}{c_W^4} \right].
\end{align}
The couplings of the $Z$ gauge boson with fermion are presented in Table \ref%
{Zcoupling},  ignoring mixing of SM and exotic quarks.
\begin{table}[ht]
	\caption{Couplings between $Z$ boson and fermions}
	\label{Zcoupling}.
	\begin{center}
		\begin{tabular}{|c|c|c|}
			\hline
			& $g_L $ & $g_R $ \\ \hline
			$\nu_i$ & $\fr{c_\phi}2 + \fr{s_\phi(-1+2 s_W^2)}{2\sqrt{3-4s^2_W}}$ & $%
			\fr{s_\phi c_W^2}{\sqrt{1-4s_W^2}}$ \\ \hline
			$e_i$ & $c_\phi \left( -\fr 1 2 + s^2_W \right) + \fr{s_\phi(-1+2 s_W^2)%
			}{2\sqrt{3-4s^2_W}}$ & $c_\phi s^2_W + s_\phi \fr{ s^2_W }{\sqrt{3-4s^2_W}}
			$ \\ \hline
			$U_n$ & $\fr{ c_\phi}6 (3-4s_W^2) + s_\phi \fr{\sqrt{3-4s^2_W}}{6} $ & $%
			- \fr 2 3 c_\phi s^2_W -\fr{2s_\phi s_W^2}{3\sqrt{3-4s_W^2}} $ \\ \hline
			$D_n$ & $\fr{ c_\phi}6 (-3+2s_W^2) + s_\phi \fr{\sqrt{3-4s^2_W}}{6} $ & $%
			\fr 1 3 c_\phi s^2_W+\fr{s_\phi s_W^2}{3\sqrt{3-4s_W^2}}$ \\ \hline
			$U_3$ & $\fr{ c_\phi}6 (3-4s_W^2) + \fr{s_\phi(-3+2 s_W^2)}{6\sqrt{%
					3-4s^2_W}} $ & $- \fr 2 3 c_\phi s^2_W - s_\phi \fr{2 s^2_W }{3\sqrt{%
					3-4s^2_W}}$ \\ \hline
			$D_3$ & $\fr{ c_\phi}6 (-3+2s_W^2) + \fr{s_\phi(-3+2 s_W^2)}{6\sqrt{%
					3-4s^2_W}}$ & $\fr 1 3 c_\phi s^2_W + s_\phi \fr{ s^2_W }{3\sqrt{3-4s^2_W%
			}}$ \\ \hline
			$T$ & $-\fr 2 3 c_\phi s^2_W - \fr{s_\phi(-3+5 s_W^2)}{3\sqrt{3-4s^2_W}}
			$ & $-\fr 2 3 c_\phi s^2_W - s_\phi \fr{2 s^2_W }{3\sqrt{3-4s^2_W}}$ \\
			\hline
			$J_n$ & $\fr 1 3 c_\phi s^2_W - s_\phi \fr{\sqrt{3-4s^2_W}}{3} $ & $%
			\fr 1 3 c_\phi s^2_W + s_\phi \fr{ s^2_W }{3\sqrt{3-4s^2_W}}$ \\ \hline
		\end{tabular}%
	\end{center}
\end{table}

It can be seen that $s_\phi \rightarrow0$ when $m_Z^2/M^2_{Z'}\rightarrow 0$, leading to the consequence that $g_L\simeq g_R$ for the exotic quarks $T,J_{1,2}$, as given in table~\ref{Zcoupling}. Note that in the limit $\phi \rightarrow 0$, the couplings of $Z$ to the SM fermions are  the same
as those of the SM $Z$ boson.

\item Interaction between $Z^\prime$ with fermion
\begin{align}
\mathcal{L}_{Z^\prime f} &= \fr{g}{c_W}\overline{f} \ga^\mu \left[ c_\phi \left(
\fr{\sqrt{3-4s^2_W}}{\sqrt{3}} T_8 + \fr{ s^2_W }{\sqrt{3-4s^2_W}}
X\right) + s_\phi \left( T_3 -s^2_W Q\right) \right] f Z^\prime_\mu \crn
&\equiv \fr{g}{c_W}\overline{f}_{L,R} \ga^\mu
g^{\prime}_{L,R} f_{L,R} Z^\prime_\mu,
\label{eq199}
\end{align}
 It is worth noting that couplings of $Z$ and $Z^\prime$ are related to each
 other by replacing $c_\phi \leftrightarrow s_\phi$.

 The couplings of the $Z_\prime$ gauge boson with fermion (by replacing $%
 c_\phi \rightarrow s_\phi $ and $s_\phi \rightarrow -c_\phi $) are presented
 in Table \ref{Zpcoupling} .
 \begin{table}[ht]
 	\caption{Couplings between $Z^\prime$ boson and fermions}
 	\label{Zpcoupling}
 	\begin{center}
 		\begin{tabular}{|c|c|c|}
 			\hline
 			& $g^\prime_L $ & $g^\prime_R $ \\ \hline
 			$\nu_i$ & $\fr{s_\phi}2 + \fr{c_\phi(-1+2 s_W^2)}{2\sqrt{3-4s^2_W}}$ & $-%
 			\fr{c_\phi c_W^2}{\sqrt{1-4s_W^2}}$ \\ \hline
 			$e_i$ & $s_\phi \left( -\fr 1 2 + s^2_W \right) - \fr{c_\phi(-1+2 s_W^2)%
 			}{2\sqrt{3-4s^2_W}}$ & $s_\phi s^2_W - c_\phi \fr{ s^2_W }{\sqrt{3-4s^2_W}}
 			$ \\ \hline
 			$U_n$ & $\fr{ s_\phi}6 (3-4s_W^2) - c_\phi \fr{\sqrt{3-4s^2_W}}{6} $ & $%
 			- \fr 2 3 s_\phi s^2_W +\fr{2c_\phi s_W^2}{3\sqrt{3-4s_W^2}} $ \\ \hline
 			$D_n$ & $\fr{ s_\phi}6 (-3+2s_W^2) - c_\phi \fr{\sqrt{3-4s^2_W}}{6} $ & $%
 			\fr 1 3 s_\phi s^2_W-\fr{c_\phi s_W^2}{3\sqrt{3 -4s_W^2}}$ \\ \hline
 			$U_3$ & $\fr{ s_\phi}6 (3-4s_W^2) - \fr{c_\phi(-3+2 s_W^2)}{6\sqrt{%
 					3-4s^2_W}} $ & $- \fr 2 3 s_\phi s^2_W + c_\phi \fr{2 s^2_W }{3\sqrt{%
 					3-4s^2_W}}$ \\ \hline
 			$D_3$ & $\fr{ s_\phi}6 (-3+2s_W^2) - \fr{c_\phi(-3+2 s_W^2)}{6\sqrt{%
 					3-4s^2_W}}$ & $\fr 1 3 s_\phi s^2_W - c_\phi \fr{ s^2_W }{3\sqrt{3-4s^2_W%
 			}}$ \\ \hline
 			$T$ & $-\fr 2 3 s_\phi s^2_W - \fr{c_\phi(3-5 s_W^2)}{3\sqrt{3-4s^2_W}}
 			$ & $-\fr 2 3 s_\phi s^2_W + c_\phi \fr{2 s^2_W }{3\sqrt{3-4s^2_W}}$ \\
 			\hline
 			$J_n$ & $\fr 1 3 s_\phi s^2_W + c_\phi \fr{\sqrt{3-4s^2_W}}{3} $ & $%
 			\fr 1 3 s_\phi s^2_W - c_\phi \fr{ s^2_W }{3\sqrt{3-4s^2_W}}$ \\ \hline
 		\end{tabular}%
 	\end{center}
 \end{table}

\eit

Note that in both Tables, dealing with neutrino we used $\nu^c_L
\sim \nu_R$.

For practical uses, we present neutral currents in the vector and axial forms as follows
\bea
\mathcal{L}_{Z f} &= & \fr{g}{2c_W}\overline{f} \ga^\mu ( g_V - \ga_5 g_A)
f Z_\mu \, ,  \label{eq931} \\
\mathcal{L}_{Z^\prime f} &= & \fr{g}{2c_W}\overline{f} \ga^\mu ( g^\prime_V -
\ga_5 g^\prime_A) f Z^\prime_\mu \, ,  \label{eq932}
\eea
where the relation among two kinds of couplings is given by
\be
g_V = g_L + g_R \, , \hspace*{0.3cm} g_A = g_L - g_R\, .  \label{eq933}
\ee

\bibliographystyle{utphys}
\bibliography{Biblio331A4October22th2021}

\end{document}